\documentclass[12pt]{article}

\usepackage{theorem,amssymb,amsmath,amsbsy,latexsym}
\usepackage{graphicx}

\newcommand{\ttpV}{$t$-$t'$-$V\,$}

\newcommand{\pdag}{^{\phantom\dag}}

\newcommand{\file}{eps}

\newcommand{\VV}{H^\prime} 

\newcommand{\EE}{\varepsilon} 

\newcommand{\fLL}{\mbox{$(\frac1{L})^2$}}
\newcommand{\QQPi}{\mbox{$\frac{2Q}{\pi}$}} 
\newcommand{\tPiL}{\mbox{$\frac{2\pi}{L}$}} 
\newcommand{\PiL}{\mbox{$\frac{\pi}{L}$}} 
\newcommand{\half}{\mbox{$\frac12$}}
\newcommand{\quart}{\mbox{$\frac14$}}

\newcommand{\tint}{\;\,\tilde{\!\!\!\int}}
\newcommand{\cL}{\mathcal{L}} 
\newcommand{\cZ}{\mathcal{Z}} 
\newcommand{\cD}{\mathcal{D}} 
\newcommand{\cE}{\mathcal{E}}

\newcommand{\ee}{\,\mathrm{e}}
\newcommand{\ii}{\mathrm{i}}

\newcommand{\BZ}{\mathrm{BZ}}
\newcommand{\C}{\mathrm{C}}

\newcommand{\Z}{{\mathbb Z}}
\newcommand{\R}{{\mathbb R}}
\newcommand{\N}{{\mathbb N}}

\newcommand{\define}{\stackrel{\mbox{{\tiny def}}}{=}}

\newcommand{\vx}{{\bf x}}
\newcommand{\vy}{{\bf y}}
\newcommand{\vn}{{\bf n}}
\newcommand{\vk}{{\bf k}}
\newcommand{\vQ}{{\bf Q}}
\newcommand{\vzero}{{\boldsymbol 0}}
\newcommand{\ve}{{\bf e}}
\newcommand{\vp}{{\bf p}}

\newcommand{\Ref}[1]{(\ref{#1})}

\begin{document}
\begin{flushright}

April 14, 2010
\end{flushright}
\vspace{.4cm}

\begin{center}

\vspace{1 cm}
{\Large \bf A 2D Luttinger model}\\
[2mm] 
{\bf Edwin Langmann} \\
[2mm] 
{\it Theoretical Physics, KTH\\
SE-10691 Stockholm, Sweden}\\ 
{\tt langmann@kth.se} \\
[5mm]
\end{center}

\begin{abstract}
  A detailed derivation of a two dimensional (2D) low energy effective
  model for spinless fermions on a square lattice with local
  interactions is given. This derivation utilizes a particular
  continuum limit that is justified by physical arguments. It is shown
  that the effective model thus obtained can be treated by exact
  bosonization methods. It is also discussed how this effective model
  can be used to obtain physical information about the corresponding
  lattice fermion system.
\end{abstract}

\section{Introduction}
\label{sec1}
One powerful approach to 1D lattice fermion systems is to utilize a
particular continuum limit to derive a low-energy effective model that
can be solved analytically. For the simplest case of spinless fermions
with local interactions away from half-filling, this approach leads to
the so-called Luttinger model \cite{Lutt} which can be solved exactly
using bosonization \cite{LM}; see \cite{T,Th,J} for closely related
pioneering papers. Exact solubility means a lot in this case: not only
the partition function but all Green's functions of the Luttinger
model can be computed exactly by analytical methods; see e.g.\
\cite{LLL,Voit} and references therein. This method can be generalized
to 1D Hubbard-like systems and is the basis of a paradigm for 1D
interacting fermion systems \cite{Haldane} (for a textbook
presentation see e.g.\ \cite{Tsvelik}, Sections~29--32).  In a
previous paper we outlined a similar approach to 2D lattice fermion
systems \cite{EL0}. The aim of this paper is to present details and
generalizations\footnote{The discussion in \cite{EL0} was restricted
  to the special parameter values $t'=0$, $\kappa=1/2$ and $\nu_a=1/2$
  (these parameters are explained in the next paragraph and after
  \Ref{Qrestr} and \Ref{nutot1}, respectively).} of this
proposal. Applications and additional results are given elsewhere
\cite{dWL1,dWL2}.

Our starting point is the so-called 2D \ttpV model describing spinless
fermions on a square lattice with nearest-neighbor and
next-nearest-neighbor hopping constants $t$ and $t'$, respectively,
and with nearest-neighbor density-density interaction strength $V/2$;
see \Ref{H0}--\Ref{V0} for a precise definition. We propose a
particular continuum limit similar to one allowing to derive the
Luttinger model from the 1D \ttpV model. Using that we derive a 2D
fermion model that is a natural 2D analogue of the Luttinger model not
only by its relation to 2D lattice fermions but also because it can be
treated by exact bosonization methods. However, different from the 1D
case, we find that only parts of the fermion degrees of freedom of
this 2D Luttinger model\footnote{The name ``2D Luttinger model'' here
  and in the following is short for ``2D analogue of the Luttinger
  model'', and we do {\em not} intend to suggest by this name that
  this model necessarily has ``Luttinger-liquid'' behavior
  \cite{Anderson1}.}  can be bosonized and thus treated exactly. We
suggest to treat the other degrees of freedom by mean field theory,
and we argue that there is a significant doping regime where this
treatment is adequate. We refer to the fermion degrees of freedom that
can be bosonized as {\em nodal} and the others as {\em antinodal}.

The 2D \ttpV model is a spinless variant of the 2D Hubbard model which
is often regarded as prototype model for high temperature
superconductors (HTSC) \cite{Anderson}. As discussed in
Section~\ref{sec3.1}, our approach and terminology is partly motivated
by the so-called pseudogap in HTSC \cite{Bonn}. Another motivation are
mean field theory results about 2D Hubbard-like models; see \cite{LW}
and references therein. Our approach was also inspired by
renormalization group studies of the 2D Hubbard model that predict a
truncation of the Fermi surface to four disconnected arcs
\cite{FRS,HSFR}; see also \cite{HM}.

Our work owes much to other ideas on 2D interacting fermion systems
discussed extensively in the literature, mainly motivated by the HTSC
problem. The idea that 2D interacting fermions can be understood by
bosonization in each direction of the Fermi surface \cite{Luther1} was
advocated by Anderson \cite{Anderson1}.  One realization of this idea
close to ours appeared already earlier: Mattis \cite{Mattis} pointed
out an exactly solvable 2D model of interacting fermions similar to
our 2D Luttinger model but without the antinodal fermions; see
\cite{Hlubina} for other work on Mattis' model. Various other
implementations of this idea have appeared in the literature since
then (see e.g.\ \cite{HKM} and references therein), but, to our
knowledge, they all differ in detail from ours. One approach similar
to ours is by Luther \cite{Luther2} who proposed an effective model
for 2D lattice fermions taking into account the 2D square Fermi
surface in all detail. In Luther's treatment an exactly solvable
bosonized model is obtained only if one ignores interactions coupling
different 1D ``chains'' in Fourier space.  Since we use a simplified
description of the Fermi surface we can avoid such a
truncation. Still, our description of the Fermi surface is more
detailed than in Mattis' model \cite{Mattis} since we also take into
account the antinodal fermions. The importance of the antinodal
fermions in 2D Hubbard-like systems is well-known from various
different approaches to the HTSC problem, including Schulz' work on
Fermi surface instabilities \cite{Schulz} and the so-called van Hove
singularity scenarios \cite{vH}.

A few remarks on our level of mathematical rigor is in order. As we
show, the 2D Luttinger model is a quantum field theory, i.e.\ a
quantum model with infinitely many degrees of freedom, that is
mathematically well-defined. However, its precise definition includes
short- and long distance regularizations whose specification requires
a considerable amount of notation. We therefore first present a formal
definition of this model in which these regularizations are
suppressed.\footnote{A similar formal definition of the
  Thirring-Luttinger model \cite{Th,Lutt} is common in the physics
  literature.}  This formal definition has heuristic value, and it
allows a concise summary of computations and results. Our derivation
of the 2D Luttinger model from the 2D \ttpV model determines the
details of all regularizations and the model parameters.  This
derivation is done by by exact computations, except for the continuum
limit that corresponds to a sequence of approximations marked as
A1--A5 in Sections~\ref{sec5.2} and \ref{sec5.4}. Physical motivations
for these approximations are discussed in Section~\ref{sec3}.  Our
treatment of the 2D Luttinger model by bosonization can be made
mathematically precise. However, to limit the length of the present
paper, we only show in Appendix~\ref{appB} that the same mathematical
results that allow to construct and bosonize the 1D Luttinger model
rigorously \cite{CR} are sufficient to do so also in the 2D case.  Our
bosonization results are otherwise derived by theoretical physics
arguments.  Mathematically precise formulations and proofs of the
latter results are given elsewhere \cite{dWL2}.

Section~\ref{sec2} contains a summary of our results, including a
formal definitions of the 2D Luttinger model and a formal description
of its bosonized form.  Section~\ref{sec3} discusses physical
motivations for our approach.\footnote{Mathematically inclined readers
  may wish to skip Sections~\ref{sec2} and \ref{sec3} at first
  reading.}  Section~\ref{sec4} fixes our notation and gives
mathematically precise definitions of the 2D \ttpV model and the 2D
Luttinger model. Section~\ref{sec5} contains a detailed derivation of
the 2D Luttinger model. Section~\ref{sec6} discusses the bosonization
of the 2D Luttinger model, the exact solution of the bosonized nodal
model, and a derivation of the effective antinodal model. We conclude
with complementary remarks in Section~\ref{sec7}. Some technical
details are deferred to three appendices.

\section{Summary of results}
\label{sec2}
We emphasize that the Hamiltonians and (anti)commutator relations of
boson and fermion operators in this section are only formal, and they
become well-defined only with particular short- and long-distance
regularizations made precise in Sections~\ref{sec4.2} and
\ref{sec5.5}.  In particle physics parlance, the Hamiltonian below
defines a quantum field theory model with UV divergences, and is is
well-defined only with a UV cutoff $a>0$ (lattice constant) and a IR
cutoff $L<\infty$ (square root of system area).  The importance of the
former cutoff is highlighted by the non-trivial scaling of the model
parameters and the boson operators with $a$; see \Ref{const}, \Ref{gj}
and \Ref{bosons} below. This scaling and the short- and long-distance
regularizations are such that physical results computed from the 2D
Luttinger model have a well-defined quantum field theory limit $L/a\to
\infty$ (this is shown for various special cases in the present paper
and in \cite{dWL1,dWL2}, and we conjecture that it is always true).

The 2D Luttinger model is formally given by the Hamiltonian $H=H_n +
H_a+H_{na}$ with
\begin{equation}
\label{Hn1} 
\begin{split} 
  H_n &= \sum_{s=\pm }\int d^2 x\, \Bigl(\sum_{r=\pm} rv_F :\!
  \psi^\dag_{r,s}(\vx)(-\ii \partial_s)\psi_{r,s}^{\phantom\dag}(\vx)\!:
  \\
  &+ g_1 \sum_{r=\pm} J_{r,s}(\vx)J_{-r,s}(\vx) + g_2
  \sum_{r,r'=\pm} J_{r,s}(\vx)J_{r',-s}(\vx) \Bigr)
\end{split}
\end{equation} 
the part involving only on the nodal fermions, 
\begin{equation}
\label{Ha1} 
H_a = \int d^2 x\,  \sum_{r=\pm}\Bigl( :\! \psi^\dag_{r,0}(\vx) 
[rc_F \partial_+\partial_- + c_F'(\partial_+^2 + \partial_-^2) 
-\mu_0 ]\psi_{r,0}^{\phantom\dag}(\vx)\!:
+ g_3 J_{r,0}(\vx)J_{-r,0}(\vx)\Bigr)
\end{equation}
the part with only antinodal fermions, and
\begin{equation}
\label{Hna1} 
H_{na} = \int d^2 x\,  
g_4\sum_{r,r',s=\pm} J_{r,0}(\vx)J_{r',s}(\vx)
\end{equation}
the mixed part.  The colons indicate normal ordering,
$\partial_{\pm}=\partial/\partial x_\pm$ with $x_\pm$ the components
of the 2D position coordinates $\vx$, $\psi_{r,s}^{(\dag)}(\vx)$ are
fermion field operators with the usual anticommutator relations $\{
\psi^{\phantom\dag}_{r,s}(\vx), \psi^{\dag}_{r',s'}(\vy)\} =
\delta_{r,r'}\delta_{s,s'}\delta^2(\vx-\vy)$ etc., and
\begin{equation}
\label{Jrs1} 
  J_{r,s}(\vx) = \, :\!  \psi^\dag_{r,s}(\vx)
  \psi_{r,s}^{\phantom\dag}(\vx)\!:
\end{equation}
are the corresponding fermion densities. The model parameters $c_F$,
$c_F'$, $v_F$, $g_j$ and $\mu_0$ are given below.

Our main result is a derivation of the 2D Luttinger model from
the 2D \ttpV model in Section~\ref{sec5} using certain 
Approximations~A1--A5. This derivation determines the regularizations
needed to make precise mathematical sense of this model. It also
introduces two parameters $\kappa$ and $Q$ in the ranges
\begin{equation}
\label{Qrestr} 
0\leq\kappa\leq 1,\quad \pi(1-\kappa)/2 < Q < \pi(1+\kappa)/2
\end{equation}
determining the size and location of the nodal Fermi surface (see
Section~\ref{sec3.2}).  This derivation also fixes the model
parameters
\begin{equation}
\label{const} 
v_F = 2\sqrt{2}\sin(Q)[t+2t'\cos(Q)]a ,\quad c_F= 2ta^2, 
\quad c_F' =  2t'a^2
\end{equation} 
\begin{equation}
\label{mua}
\mu_0 = -4t' -[4t+V(1-\kappa)^2]\cos(Q) -
[4t'+2V(1-\kappa)(\QQPi-1)]\cos^2(Q) 
\end{equation} 
\begin{equation}
\label{gj}
g_1=2g_2 = 2V a^2 \sin^2(Q),\quad g_3=g_4 = 2V a^2
\end{equation}
and the filling factor
\begin{equation}
\label{nutot1} 
\nu = \half + (1-\kappa)\left(\QQPi-1\right)+\kappa^2(\nu_a-\half). 
\end{equation} 
The parameter $\nu_a$ is the filling factor of the anti-nodal fermions
and can be computed self-consistently \cite{dWL1}. 

Our second result is to show that the 2D Luttinger model is equivalent
to a model of 2D bosons coupled to the antinodal fermions and with
interactions at most quadratic in the bosons field operators
(Section~\ref{sec6.1}). A mathematically precise statement of this
result is given in proposition~\ref{prop1}. It can be formally
summarized as follows: the following linear combinations of fermion
densities
\begin{equation}
\label{bosons}
\begin{split}
  \partial_s\Phi_s(\vx)&= \sqrt{\frac{\tilde{a}}{4\pi}}[J_{+,s}(\vx) +
  J_{-,s}(\vx)] \\
  \Pi_s(\vx)&= \sqrt{\frac{\tilde{a}}{4\pi}}[-J_{+,s}(\vx) +
  \hat{J}_{-,s}(\vx)]
\end{split} 
\end{equation} 
with 
\begin{equation}
\label{tildea}
\tilde{a}\define  \frac{\sqrt{2}}{1-\kappa} a
\end{equation} 
define 2D boson operators obeying the canonical commutator relations
$[\Pi\pdag_s(\vx),\Phi_{s'}(\vy)] =-\ii\delta_{s,s'}\delta^2(\vx-\vy)$
etc. Moreover, the nodal and mixed parts of the 2D Luttinger
Hamiltonian are equal to
\begin{equation}
\label{Hn11} 
H_n = \frac{v_F}2 \int d^2x\,:\! \Bigl(
 \sum_{s=\pm} \Bigl[(1-\gamma)\Pi_s(\vx)^2 +(1+\gamma)\bigl( 
 \partial_s\Phi_s(\vx)\bigr)^2\Bigr] +2\gamma\partial_+\Phi_+(\vx) 
 \partial_-\Phi_-(\vx)
 \Bigr)\!:   
\end{equation}
and 
\begin{equation}
\label{Hna11} 
H_{na} = \int d^2 x\, \frac{g_4}{\sqrt{\pi\tilde{a}}}
\sum_{r,s=\pm}  J_{r,0}(\vx)\partial_s\Phi_s(\vx) 
\end{equation} 
with the dimension-less coupling parameter 
\begin{equation}
  \gamma =   \frac{V(1-\kappa)\sin(Q)}{2\pi [t+2t'\cos(Q)]}. 
  \label{gamma} 
\end{equation} 
As we will see, this implies that the 2D Luttinger model is
well-defined provided that $\gamma<1$.  Fortunately this does not
restrict the model to weak coupling.  We also outline how to exactly
solve the bosonized nodal model (Section~\ref{sec6.2}).

Our third result is an effective model for the antinodal fermions
obtained by integrating out the bosons exactly
(Section~\ref{sec6.3}). This effective antinodal model includes an
effective interaction that has a non-trivial time dependence. We
propose to approximate this interaction by an instantaneous one (as
will be discussed, this approximation is convenient but not
essential). We find that the effective antinodal model can then be
represented by an Hamiltonian as in \Ref{Ha} but with $g_3$ changed to
$g_3-g_{{\textrm{eff}}}$ with
\begin{equation}
\label{geff} 
  g_{\textrm{eff}} =  
  \frac{V^2(1-\kappa)a^2}{ \sin(Q)\pi[t + 2t'\cos(Q)
    + \frac{V}{\pi}(1-\kappa)\sin(Q)]} 
\end{equation}
the renormalization of the antinodal coupling by the nodal fermions.
Note that all parameters in this effective antinodal model scale like
$a^2$.  Mean field results for the effective antinodal model are given
in \cite{dWL1}. Green's functions of the nodal model defined by the
Hamiltonian in \Ref{Hn} can be computed exactly \cite{dWL2}.

\section{Physical motivations}
\label{sec3}
\subsection{Experiments}
\label{sec3.1} 
We recall experimental facts about HTSC \cite{Bonn} that motivate our
approach and the terminology we use.

\begin{figure}[ht!]
   \vspace{0.5in}
\begin{center}
\includegraphics[width=0.8\textwidth]{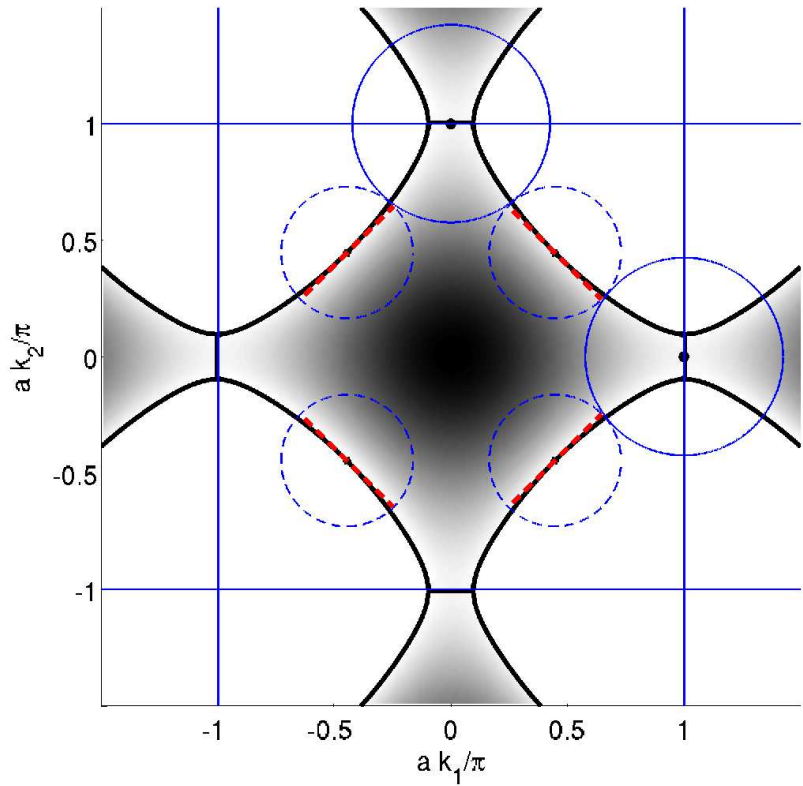}
\end{center}
\caption{Fermi surface $\epsilon(\vk)=\mu$ for the tight-binding
  dispersion relations in \Ref{eps0} for $t'/t=-0.2$ and $\mu/t
  =-0.672$. The full and dashed circles indicate the antinodal
  respectively nodal regions, as explained in the main text. The
  dashed lines indicate the Fermi surface arcs we postulate to exist
  in our approach.}
\label{FIG1}
\end{figure}

An important parameter determining the physical behavior of HTSC
materials is the so-called {\em filling factor} $\nu$ defined as the
total fermion number divided by the largest possible number of
fermions (note that our definition is such that $0\leq\nu\leq 1$).
HTSC are typically insulators at half-filling ($\nu=1/2$), and they
have a conventional and essentially 2D Fermi surface in the so-called
overdoped regime where $\nu$ is significantly different from $1/2$. In
the latter regime HTSC behave like conventional metals. Our interest
is in the so-called underdoped regime where $\nu$ is close to but
different from $1/2$. In this regime HTSC have exotic properties not
fully understood. In particular, the Fermi surface in this regime
seems to be partially gapped \cite{ARPES}: there still exists a
so-called {\em underlying Fermi surface} that can be well described by
a tight-binding dispersion relation \cite{Y1} (an example is given in
figure~\ref{FIG1}). The visible part of this Fermi surface consists
typically of four line segments called {\em nodal arcs} (enclosed by
dashed circles in figure~\ref{FIG1}), and the remaining so-called {\em
  antinodal} parts of the Fermi surface (enclosed by full circles in
figure~\ref{FIG1}) are gapped. The $\nu$-dependence of the size and
location of the nodal arcs has been measured for various HTSC
materials; see e.g.\ \cite{Y1} and references therein.

\subsection{Theory} 
\label{sec3.2} 
The continuum limit we use to derive the 2D Luttinger model is based
on the following hypothesis about interacting fermion systems that can
be justified by renormalization group arguments; see e.g.\ \cite{M},
Chapter~12 and 14 or \cite{Salmhofer}, Chapter~4.

\begin{itemize}
\item[\textbf{H1:}] \textit{There exists some underlying Fermi surface
    dominating the low energy physics of the interacting fermion
    model, and one can modify, ignore or add degrees of freedom far
    away from this Fermi surface without changing the low energy
    properties of the model much.}
\end{itemize} 

As a motivation we first recall the well-known 1D case
\cite{Haldane}. The non-interacting Fermi surface of the 1D \ttpV
model consists of two points, and according to Hypothesis~H1 one can
perform a continuum limit that amounts to the following
approximations: linearizing the dispersion relation in the vicinity of
the Fermi surface points, modifying interactions of degrees of freedom
away from these points, and partly removing the short-distance
cutoff.\footnote{The remaining short-distance regularization
  corresponds to the non-locality of the two-body interaction of the
  Luttinger model. In the limit where this interaction becomes local,
  the Luttinger model becomes equal to the massless Thirring model
  \cite{T} which is more complicated from a mathematical point of
  view; see e.g.\ \cite{GLR} for a discussion of this limit.} One thus
obtains a continuum model with two fermion branches, the left- and the
right movers, representing the fermion degrees of freedom in vicinity
of the two Fermi-surface points. The approximations described above
provide crucial simplifications since they lead to a model of
interacting fermions that can be mapped exactly to a non-interacting
boson model \cite{Haldane}.

In 2D the situation is more involved.  However, it is still useful to
consider a typical non-interacting Fermi surface of the 2D \ttpV
model.  The latter is defined by the following tight-binding
dispersion relation
\begin{equation}
\label{eps0} 
\epsilon(\vk) = -2t [\cos(ak_1) + \cos(ak_2) ] - 
4t'\cos(ak_1)\cos(ak_2)
\end{equation}
with $\vk=(k_1,k_2)$ the usual momenta in the Brillouin zone
$-\pi/a\leq k_j<\pi/a$ and the hopping parameters $t>0$ and $t'$
($a>0$ is the lattice constant). As a representative example we plot
in figure~\ref{FIG1} the Fermi surface $\epsilon(\vk)=\mu$ for some
small negative values of $t'/t$ and the Fermi energy $\mu$.

The behavior of the dispersion relation $\epsilon(\vk)$ is
qualitatively different in different regions of the Brillouin zone:
since $(\pi/a,0)$ and $(0,\pi/a)$ are saddle points of
$\epsilon(\vk)$, no linear approximation of the dispersion relation
exists in the so-called antinodal regions close to these saddle
points. However, in the so-called nodal regions close to points
$\vk=(Q/a,\pm Q/a)$ and $(-Q/a,\mp Q/a)$ for some parameter $Q\approx
\pi/2$, one can approximate by a linear dispersion relation; see
\Ref{Era}. (The antinodal and nodal regions are indicated by full
respectively dashed circles in figure~\ref{FIG1}.)  This suggests that
the fermion degrees of freedom in these different regions play
different roles in the interacting model.  We thus rewrite the model
by introducing field operators $\psi^{(\dag)}_{r,s}$ labeled by
``flavor indices'' $r=\pm$ and $s=0,\pm$ that represent the fermion
degrees of freedom in these different regions: $r=\pm$ and $s=\pm$
correspond to the regions close to the nodal points $(rQ/a,s rQ/a)$,
and $r=\pm$ and $s= 0$ correspond to the regions close to the
antinodal points $(\pi/a,0)$ and $(0,\pi/a)$, respectively. Before we
can use H1 above to modify the model we need another hypothesis that
is partly motivated by experiments on HTSC mentioned in
Section~\ref{sec3.1}:

\begin{itemize}
\item[\textbf{H2:}] \textit{The underlying Fermi surface contains
    nodal arcs that can be well approximated by straight line
    segments.}
\end{itemize}

To be more specific, we assume that the Fermi surface contains four
points $(Q/a,\pm Q/a)$ and $(-Q/a,\mp Q/a)$ for some parameter $Q
\approx \pi/2$, and that close to each of these points we can
approximate the exact dispersion relation by the linearized one. As
will be seen, this corresponds to a Fermi surface containing four
straight line segments as indicated by the dashed lines in
figure~\ref{FIG1}. We introduce a parameter $\kappa$ determining the
size of the regions where we use the linearized dispersion relations
and such that the length of these line segments is
$(1-\kappa)\sqrt{2}\pi/a$. We emphasize that we do not make any
assumption about the underlying Fermi surface in the antinodal region,
and whether the antinodal regions are gapped or not depends on
parameters and can be determined by computations \cite{dWL1}.

Our approximations leading to the 2D Luttinger model are similar to
the ones in 1D mentioned above. They are (essentially) justified by
Hypotheses~H1 and H2 above. Similarly as in 1D, due to their linear
dispersion relation, the interacting nodal fermions can be mapped
exactly to non-interacting bosons.

\section{Definitions and notation}
\label{sec4}
We write ``$\define$'' to emphasize that an equality is a definition.
We write a vector $\vk\in \R^2$ in components either as
$\vk=(k_1,k_2)$, $k_j\in\R$, or as $\vk=k_+\ve_++k_-\ve_-$ with
\begin{equation*}
  k_\pm \define (k_1\pm k_2)/\sqrt{2},\quad 
  \ve_\pm \define (1,\pm 1)/\sqrt{2}, 
\end{equation*}
and $|\vk|\define\sqrt{k_+^2+k_-^2}$. We use symbols $\vx,\vy$ for
fermion positions, $\vk,\vk'$ for fermion momenta, and $\vp$ for
differences of fermion momenta. 

We write ``$k_\pm< c$'' short for ``$k_+<c$ and $k_-< c$'' etc., and
``$k_\pm\not < c$'' is short for ``$k_+\not < c$ or $k_-\not< c$ (or
both)'' etc.

\subsection{2D \ttpV model}
\label{sec4.1}
We consider a diagonal square lattice $\Lambda$ with lattice constant
$a>0$ and $(L/a)^2\gg 1$ lattice sites: The lattice sites are
$\vx=(x_1,x_2)=(an_1,a n_2)$ with integers $n_j$ such that $-L/2\leq
x_\pm < L/2$, and we assume that $L/(2\sqrt{2}a)\in\N$.\footnote{We
  this somewhat unconventional large-distance regularization to
  simplify some technicalities later on.} The momenta we use are
therefore restricted to be in the following sets,
\begin{equation}
\begin{split}
  \Lambda^*
  \define& \left\{ \vk\in\R^2 :\;  k_\pm\in\tPiL(\Z+\half)\right\}  \\
  \tilde\Lambda^* \define& \left\{ \vp\in\R^2 :\; p_\pm\in\tPiL \Z
  \right\}
\end{split} 
\end{equation} 
(we use antiperiodic boundary conditions for the fermions).  The
Brillouin zone (lattice Fourier space) for the fermions is
\begin{equation}
  \label{BZ} 
  \BZ \define \left\{ \vk \in\Lambda^* :\; 
   -\frac{\pi} a \leq k_j <\frac{\pi} a,\quad j=1,2 \right\}. 
\end{equation} 

The 2D \ttpV model is defined on a Fock space with fermion creation-
and annihilation operators $\psi^{(\dag)}(\vx)$ labeled by lattice
vectors $\vx$ and satisfying the usual canonical anticommutator
relations. We denote as $|0\rangle$ the normalized vacuum state
annihilated by all $\psi(\vx)$, and our normalization is such that $\{
\psi(\vx), \psi^\dag(\vy)\} = \delta_{\vx,\vy}/a^2$. The Hamiltonian
of this model is
\begin{equation}
\label{HttpV}
H_{tt'V}=H_0+\VV
\end{equation}
with the free part 
\begin{equation}
\label{H0}
H_0 = \sum_{\vx,\vy\in\Lambda}a^4 [-T(\vx-\vy)- \mu 
\delta_{\vx,\vy}/a^2 ]\psi^\dag(\vx)\psi(\vy)
\end{equation}
and the hopping matrix $T(\vx-\vy)$ equal to $t/a^2>0$ for nearest
neighbor sites (i.e.\ if $|\vx-\vy|=a$), $t'/a^2$ for next-nearest
neighbor sites (i.e.\ if $|\vx-\vy|=\sqrt 2 a$), and zero otherwise;
$\mu\in\R$ is the chemical potential. The interaction part is
\begin{equation}
\label{V0} 
  \VV = \sum_{\vx,\vy\in\Lambda} a^4 u(\vx-\vy)\psi^\dag(\vx)\psi(\vx) 
  \psi^\dag(\vy)\psi(\vy) 
\end{equation} 
with $u(\vx-\vy)=V/4>0$ for nearest neighbor sites $\vx$, $\vy$ and
zero otherwise. Note that our scaling with $a$ is such that $t$, $t'$,
$V$ and $\mu$ all have the dimensions of an energy. We assume that
$|2t'/t|<1$ (otherwise the qualitative features of the dispersion
relation are different from what we assume in Section~\ref{sec5} and
our treatment does not apply).

The average number of fermions per site is the filling factor (or
filling) defined as follows,
\begin{equation}
\label{nudef}
\nu\define \left( \mbox{$\frac{a}{L}$} \right)^2 \langle N\rangle,
\quad N\define \sum_{\vx\in\Lambda}a^2\psi^\dag(\vx)\psi(\vx)
\end{equation}
where $\langle\cdot\rangle$ is the ground (or thermal) state
expectation value and $N$ the fermion number operator. Thus $0\leq
\nu\leq 1$, and half-filling corresponds to $\nu=1/2$. We sometimes
refer to $\nu-1/2$ as {\em doping}.

We recall that the model is invariant under the particle-hole
transformation $\psi(\vx)\leftrightarrow (-
1)^{(x_1+x_2)/a}\psi^\dag(\vx)$ equivalent to the following change of
parameters
\begin{equation}
  \label{ParticleHole}
(t,t',\nu,\mu,V) \to (t,-t',1-\nu,2V-\mu,V). 
\end{equation}
The model is also invariant under the following (discrete) rotation
and parity transformations
\begin{equation}
\label{RP}
\mathcal{R}: \; \vx=(x_1,x_2)\to (-x_2,x_1),\quad \mathcal{P}:\; \vx\to -\vx. 
\end{equation} 

Our conventions for Fourier transformation are as follows,
\begin{equation*}
\hat\psi(\vk) = \frac1{2\pi} \sum_{\vx\in\Lambda} a^2
\psi(\vx)\ee^{-\ii\vk\cdot\vx} 
\end{equation*}
with $\vk\in\BZ$. This allows us to write
\begin{equation}
\label{hH0}
  H_0 = \sum_{\vk\in \BZ} (\tPiL)^2 [\epsilon(\vk)-\mu] 
  \hat\psi^\dag(\vk)\hat\psi(\vk) 
\end{equation}
with the dispersion relation in \Ref{eps0}. The interaction part in
Fourier space is
\begin{equation}
\label{hV}
\VV = \sum_{\vk_j\in\BZ} \left(\tPiL\right)^{8}
\hat{v}(\vk_1,\vk_2,\vk_2,\vk_4)
\, \hat\psi^\dag(\vk_1)\hat\psi(\vk_2)
\hat\psi^\dag(\vk_3)\hat\psi(\vk_4)  
\end{equation}  
with 
\begin{equation}
\label{hatv}
\hat{v}(\vk_1,\vk_2,\vk_3,\vk_4) = \hat{u}(\vk_1-\vk_2) 
\sum_{\vn\in\Z^2} \left(\mbox{$\frac{L}{2\pi}$} \right)^2 
\delta_{\vk_1-\vk_2+\vk_3-\vk_4,2\pi\vn/a}     
\end{equation} 
and 
\begin{equation}
\label{hatu} 
\hat{u}(\vp)=\frac{a^2 V}{8\pi^2} [\cos(a p_1)+\cos(a p_2)]. 
\end{equation} 
The physical interpretation of our interaction is that it contains all
possible scattering terms $\vk_4\to\vk_3$ and $\vk_2\to\vk_1$ weighted
by a factor $\hat{u}(\vk_1-\vk_2)$ and otherwise restricted only by
overall momentum conservation up to umklapp processes.

We also use fermion density operators 
\begin{equation*}
\rho(\vx) \define \psi^\dag(\vx)\psi(\vx)
\end{equation*}
with the following conventions for Fourier transformation
\begin{equation*}
\hat\rho(\vp) = \sum_{\vx\in\Lambda} a^2 \rho(\vx)\ee^{-\ii \vp\cdot\vx}
= \sum_{\vk_1,\vk_2\in\BZ}  \hat\psi^\dag(\vk_1)\hat\psi(\vk_2)
\sum_{\vn\in\Z^2}\delta_{\vk_1+\vp,\vk_2+2\pi\vn/a}
\end{equation*}
for $\vp\in\tilde\Lambda^*$ such that $|p_\pm|\leq\pi/a$. This
allows us to write
\begin{equation*}
\VV=\sum_{\vx,\vy}a^4 u(\vx-\vy)\rho(\vx)\rho(\vy)
 = \sum_{\vp}\fLL \hat{u}(\vp)\hat\rho(-\vp)\hat\rho(\vp).  
\end{equation*}
Our conventions for density operators of the 2D Luttinger model are
similar.

\noindent \textbf{Remark:} Note that our normalizations are such that
the naive continuum limit $a\to 0$ makes sense, in particular
$\delta_{\vx,\vy}/a^2\to \delta^2(\vx-\vy)$ (Dirac delta) and
$\sum_{\vx} a^2\to \int d^2 x$ (Riemann sum). This limit corresponds
to the one discussed in Section~\ref{sec7}, remark 4.  However, we are
mainly interested in the continuum limit where $\nu$ is fixed and
close to $1/2$, and this limit is more delicate.  Note also that our
formulas have a well-defined limit $L\to\infty$, in particular,
$\{\hat\psi(\vk),\hat\psi^\dag(\vk')\} =
[L/(2\pi)]^2\delta_{\vk,\vk'}\to\delta^2(\vk-\vk')$ and
$\sum_{\vk}(2\pi/L)^2\to \int d^2k$.

\subsection{2D Luttinger model}
\label{sec4.2}
The formal definition of the 2D Luttinger model in Section~\ref{sec2}
can be made mathematically precise as follows.

The 2D Luttinger model is an effective model for the 2D \ttpV model
depending on two more parameters $\kappa$ and $Q$ in the ranges given in 
\Ref{Qrestr} and such that 
\begin{equation}
\kappa\in \frac{2\sqrt{2}a}{L}(\N+\half),\quad  Q \in\frac{\sqrt{2}\pi a}{L}\N
\label{Qquant}
\end{equation}
(it would be easy to drop the restrictions in \Ref{Qquant}, but they
simplify some formulas below, and they become irrelevant for $L/a\gg
1$).

We introduce fermion operators $\hat\psi^{(\dag)}_{r,s}(\vk)$ labeled
by two indices $s=0,\pm$ and $r=\pm$ and by momenta $\vk$ in different
Fourier spaces $\Lambda^*_s$ as follows,
\begin{equation}
  \label{Lambdas}
\begin{split}
  \Lambda^*_{0} =& \left\{ \vk\in\Lambda^*:\; |k_\pm+\PiL|<
    \frac{\kappa\pi}{\sqrt{2}a} \right\} \\
  \Lambda^*_{s=\pm} =& \left\{\vk\in\Lambda^* :\; |k_{-s}+\PiL|<
    \frac{(1-\kappa)\pi}{\sqrt{2}a}\right\}.
\end{split} 
\end{equation}
Note that there are infinitely many nodal (i.e.\ $s=\pm$) fermion
degrees of freedom (since there is no restriction in
$\Lambda^*_{s=\pm}$ on $k_s$), but only finitely many antinodal (i.e.\
$s=0$) ones. These fermion operators obey the usual canonical
anticommutator relations normalized such that
\begin{equation}
\{\hat\psi\pdag_{r,s}(\vk),\hat\psi^\dag_{r',s'}(\vk')\} 
= \delta_{r,r'}\delta_{s,s'}\delta_{\vk,\vk'}(\tPiL)^2, 
\end{equation}
and they are defined on a Fock space with a non-trivial vacuum
$\Omega$ (``Dirac sea'') characterized by the following conditions,
\begin{equation}
\label{DiracSea}
\hat\psi\pdag_{r,s=\pm}(\vk)\Omega = \hat\psi^\dag_{r,s=\pm}(-\vk)\Omega=0
\quad \forall \vk\,\mbox{ such that }\, rk_s>0
\end{equation} 
and $\langle\Omega,\Omega\rangle=1$, with $\langle\cdot,\cdot\rangle$
the Fock space inner product.  For the antinodal fermions we only
assume that antinodal filling is the same for $r=+$ and $r=-$, i.e.\
\begin{equation}
\label{nua} 
\nu_a \define \frac2{(\kappa L)^2} \sum_{\vk\in\Lambda^*_0}
(\tPiL)^2 \langle\Omega,
\hat\psi^\dag_{r,0}(\vk)\hat\psi\pdag_{r,0}(\vk)\Omega\rangle
\end{equation}
is independent of $r=\pm$. The latter assumption could be easily
dropped, but it simplifies some formulas, and it can be checked at the
end of our computations. Note that $0\leq \nu_a\leq 1$, i.e.\ $\nu_a$
defined in \Ref{nua} is the filling factor for the antinodal
fermions. It is not a free parameter but can be computed
self-consistently \cite{dWL1}. We denote by colons normal ordering
with respect to $\Omega$ as usual,
\begin{equation}
\label{NO} 
:\!A\!:\; \define A - \langle\Omega,A\Omega\rangle
\end{equation}
for fermion bilinear operators $A=\hat\psi^\dag_{r,s}(\vk)
\hat\psi\pdag_{r,s}(\vk')$ etc.\ (boson operators are normal ordered
in the same way in Section~\ref{sec6}).  We also introduce the cutoff
functions
\begin{equation}
\label{chis}
\chi_{s=\pm}(\vp)=\begin{cases}
1 & \mbox{ if }\; |p_s|\leq \frac{\kappa\pi}{\sqrt{2}a}
  \; \mbox{ and }\;  |p_{-s}|\leq \frac{(1-\kappa)\pi}{\sqrt{2}a}\\
0 & \mbox{ otherwise.} 
\end{cases}  
\end{equation}

The 2D Luttinger model is defined by the Hamiltonian
$H=H_n+H_a+H_{na}$ with
\begin{equation}
\label{Hn}
\begin{split}
  H_n =& \sum_{s=\pm} \sum_{\vk\in\Lambda^*_s}(\tPiL)^2\sum_{r=\pm} rv_F
  k_{s}
  :\!\hat\psi^\dag_{r,s}(\vk)\hat\psi^{\phantom\dag}_{r,s}(\vk)\!: 
\\  &
+ 2\sum_{\vp\in\tilde\Lambda^*}\fLL \Bigl(
  g_1\sum_{s=\pm}\chi_s(\vp) \hat{J}_{+,s}(-\vp)\hat{J}_{-,s}(\vp) +
  g_2 \sum_{r,r'=\pm}\chi_+(\vp)\chi_-(\vp)
  \hat{J}_{r,+}(-\vp)\hat{J}_{r',-}(\vp) \Bigr)
\end{split}
\end{equation} 
the nodal part, 
\begin{equation}
\begin{split}
\label{Ha}
H_a =& \sum_{\vk\Lambda^*_0}(\tPiL)^2\sum_{r=\pm} [ - rc_F k_+k_--
c_F'(k_+^2+k_-^2)-\mu_0]
:\!\hat\psi^\dag_{r,0}(\vk)\hat\psi^{\phantom\dag}_{r,0}(\vk)\!:
\\
&+ 2\sum_{\vp\in\tilde\Lambda^*}\fLL \, g_3 \hat{J}_{+,0}(-\vp)\hat{J}_{-,0}(\vp)
\end{split}
\end{equation} 
the antinodal part, and
\begin{equation}
\label{Hna}
H_{na} =  \sum_{\vp\in\tilde{\Lambda}^*} \fLL \,  g_4 \sum_{r,r'}  
\hat{J}_{r,0}(-\vp)\sum_{s=\pm}\chi_s(\vp)\hat{J}_{r',s}(\vp)
\end{equation}
the mixed part. The interactions involve the following Fourier
transformed and normal ordered fermion densities,
\begin{equation}
  \label{hatJrs}
\begin{split}
  \hat{J}_{r,0}(\vp) =& \sum_{\vk\in\Lambda^*_0}(\tPiL)^2
  :\! \hat\psi^\dag_{r,0}(\vk-\vp)\hat\psi\pdag_{r,0}(\vk)\!: \\
  \hat{J}_{r,s=\pm}(\vp) =& \sum_{\vk\in\Lambda^*_s}(\tPiL)^2 \sum_{n\in\Z}
  :\!
  \hat\psi^\dag_{r,s}(\vk-\vp)\hat\psi\pdag_{r,s}(\vk+\sqrt{2}(1-\kappa)\pi
  n\ve_{-s}/a)\!:.
\end{split}
\end{equation}
The model parameters are given in \Ref{const}--\Ref{nutot1}. As
mentioned, the 2D Luttinger model is well-defined if $\gamma$ in
\Ref{gamma} satisfies the condition $\gamma<1$, and we always assume
this is the case.

As discussed in more detail below, normal ordering is essential to
make the terms involving nodal fermions well-defined, but for the
antinodal fermions it only amounts to subtracting finite
constants. Note that $\hat{J}_{r,0}(\vp)$ is non-zero only for
$|p_\pm|\leq \sqrt{2}\pi\kappa/a$, and $\hat{J}_{r,s=\pm}(\vp)$ is
defined for $\vp$ in
\begin{equation}
\label{Lams}
\C_{s=\pm}=\left\{ \vp\in\tilde\Lambda^*\, : \, |p_{-s}|\leq 
  \frac{\pi(1-\kappa)}{\sqrt{2}a} \right\}.  
\end{equation}

The relation between the precise definition of the 2D Luttinger model
here and the formal one in Section~\ref{sec2} is spelled out in
Section~\ref{sec5.5}.

\section{Derivation of the 2D Luttinger model}
\label{sec5}

\subsection{Eight flavor model}
\label{sec5.1} 
\begin{figure}[ht!]
\begin{center}
\includegraphics[width=0.8\textwidth]{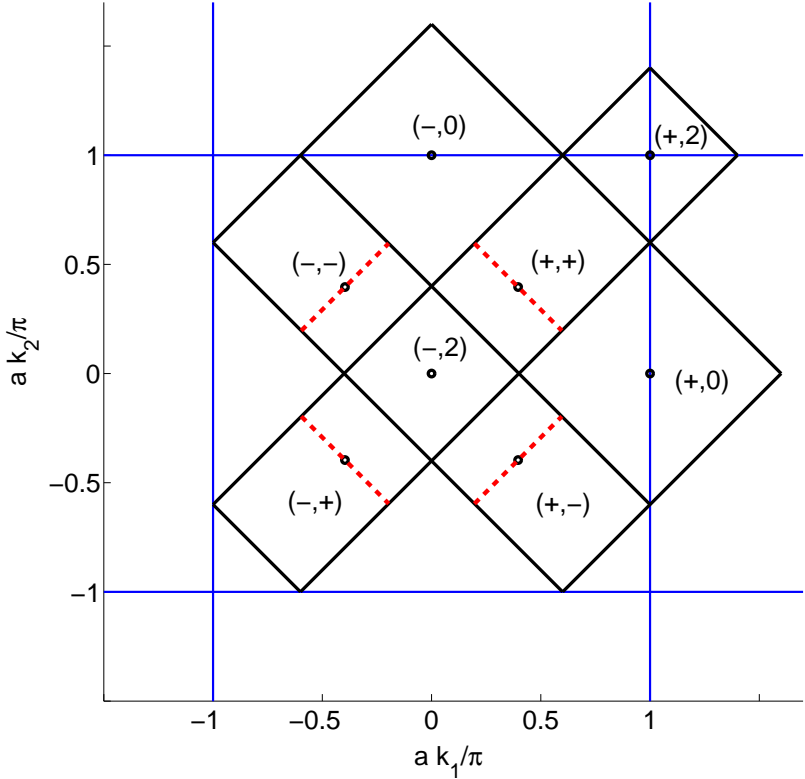}
\end{center}
\caption{Division of the Brillouin zone $\BZ$ in eight regions
  $\vQ_{r,s}+\Lambda^*_{r,s}$ marked as $(r,s)$ for $r=\pm$, $s= 0,\pm
  ,2$; see \Ref{vQ} and \Ref{Lambdars}. The eight dots mark the points
  $\vQ_{r,s}$, and the parameters are $Q=0.45\pi$ and $\kappa=0.8$.
  Note that $k_{1,2}$ can be shifted by integer multiples of $2\pi/a$
  so that the union of these regions cover the $\BZ$ exactly once.}
\label{FIG3}
\end{figure}

We divide our Brillouin zone in eight non-overlapping regions as
indicated in figure~\ref{FIG3}. For that we define eight vectors
$\vQ_{r,s}$ with $r=\pm$ and $s=0,\pm ,2$ as follows
\begin{equation}
\label{vQ}
\begin{split}
&\vQ_{+,0}=(\pi/a ,0),\quad \vQ_{-,0}=(0,\pi/a) \\
&\vQ_{r,s}= (rQ/a,rs Q/a)\; \mbox{ for } \; r=\pm,s=\pm \\
&\vQ_{-,2}=(0,0),\quad \vQ_{+,2}=(\pi/a,\pi/a)
\end{split} 
\end{equation}
for some $Q\approx \pi/2$ obeying the second condition in
\Ref{Qrestr}.\footnote{This condition ensures that
  $\vQ_{r,s=\pm}\in\tilde\Lambda^*$.} We also introduce eight
corresponding rectangular regions $\Lambda^*_{r,s}$ so that every
vector in $\BZ$ can be written uniquely as $\vQ_{r,s}+\vk+2\pi\vn/a$
for some $r\in\{+,-\}$, $s\in\{0,\pm,2\}$, $\vk\in\Lambda^*_{r,s}$,
and $\vn\in\Z^2$ (the umklapp term $2\pi\vn/a$ is sometimes needed to
translate to the region $-\pi/a\leq k_j< \pi/a$). Mathematically these
regions can be defined as follows
\begin{equation}
\label{Lambdars}
\begin{split}
  \Lambda^*_{r,0} =& \left\{ \vk\in\BZ :\; 
    |k_\pm+\PiL|<
    \frac{\kappa\pi}{\sqrt{2}a} \right\} \\
  \Lambda^*_{r,s=\pm} =& \left\{\vk\in\BZ :\; 
    \left|k_s +\PiL + r\frac{(2Q-\pi)}{\sqrt{2}a}\right| < \frac{\kappa
      \pi}{\sqrt{2}a},\quad 
    \left|k_{-s}+\PiL\right|< \frac{(1-\kappa)\pi}{\sqrt{2}a}
  \right\} \\
  \Lambda^*_{r,2} =& \left\{\vk\in\BZ:\;\left|k_\pm+\PiL\right|<
    \frac{(1-\kappa)\pi}{\sqrt{2}a }\right\} 
\end{split} 
\end{equation}
for some parameter $\kappa$ satisfying the first condition in
\Ref{Qrestr}\footnote{This condition ensures that
  $\kappa\pi/(\sqrt{2}a)$ and $(1-\kappa)\pi/(\sqrt{2}a)$ are
  half-integers multiples of $2\pi/L$.}  (see figure~\ref{FIG3}). Thus
our division of the Brillouin zone depends on two parameters $Q$ and
$\kappa$ where the former specifies the locations and the latter the
size of the Fermi surface arcs. It is easy to see that, for geometric
reasons, $\kappa$ and $Q$ are restricted as in \Ref{Qrestr}. Note that
$\Lambda^*_{r,0}$ is the same for $r=+$ and $r=-$ and equal to
$\Lambda^*_0$ in \Ref{Lambdas}.

We define
\begin{equation*}
  \hat\psi_{r,s}(\vk)\define \hat\psi(\vQ_{r,s} +\vk +2\pi\vn/a)\quad 
  \forall \vk\in\Lambda^*_{r,s}
\end{equation*}
with $\vn\in\Z^2$ such that $\vQ_{r,s} +\vk +2\pi\vn/a\in\BZ$. This
allows us to rewrite the free- and interaction parts of our
Hamiltonian in \Ref{hH0} and \Ref{hV} as follows,
\begin{equation*}
H_0 = \sum_{r,s} \sum_{\vk\in\Lambda^*_{r,s}} (\tPiL)^2 
[\epsilon(\vQ_{r,s}+\vk)-\mu]  \hat\psi^\dag_{r,s}(\vk) 
\hat\psi^{\phantom\dag}_{r,s}(\vk)
\end{equation*}
and 
\begin{equation*}
\VV = \sum_{r_j,s_j} \sum_{\vk_j\in\Lambda^*_{r_j,s_j}} \left(\tPiL\right)^{8}
\hat{v}(K_1,K_2,K_3,K_4)
\, \hat\psi^\dag_{r_1,s_1}(\vk_1) \hat\psi^{\phantom\dag}_{r_2,s_2}(\vk_2)
\hat\psi^\dag_{r_3,s_3}(\vk_3) \hat\psi^{\phantom\dag}_{r_4,s_4}(\vk_4)  
\end{equation*} 
where $K_j$ is short for $(\vk_j,r_j,s_j)$ and
\begin{equation}
\label{vv8}
\begin{split}
\hat{v}(K_1,K_2,K_3,K_4) =&
\hat{u}(\vQ_{r_1,s_1} -\vQ_{r_2,s_2} + \vk_1-\vk_2)
\sum_{\vn\in\Z^2} \left(\mbox{$\frac{L}{2\pi}$}\right)^2
\\ 
&\times \delta_{\vQ_{r_1,s_1}-\vQ_{r_2,s_2}+ \vQ_{r_3,s_3}
  -\vQ_{r_4,s_4} + \vk_1-\vk_2+\vk_3-\vk_4,2\pi /a\vn}. 
\end{split} 
\end{equation} 
The fermion number operator can now be written as
\begin{equation}
\label{Nrs} 
N = \sum_{\vk\in\BZ}(\tPiL)^2 \hat\psi^\dag(\vk)\hat\psi(\vk)=
\sum_{r,s}N_{r,s},\quad 
N_{r,s} = \sum_{\vk\in\Lambda^*_{r,s}}(\tPiL)^2 \hat\psi^\dag_{r,s}(\vk) 
\hat\psi^{\phantom\dag}_{r,s}(\vk)
\end{equation}
where $N_{r,s}$ is the number operator for the $(r,s)$ fermions. 

Note that, up to now, we only rewrote the lattice Hamiltonian without
any approximation. However, we now can interpret it as model of eight
different flavors of fermions distinguished by the labels $s$ and
$r$. The degrees of freedom with the flavor index $s=0$, $r=\pm$ are
what we call {\em antinodal fermions}, and the ones with $s=\pm$,
$r=\pm$ are the {\em nodal fermions}. We call the degrees of freedom
with $s=2$ for $r=-$ and $r=+$ the {\em in}- and the {\em out
  fermions}, respectively.

We have introduced two parameters $Q$ and $\kappa$ for generality, and
the effective model we derive below depends on them. In principle
these parameters can be determined from filling and by minimizing the
total energy, but for now we leave them arbitrary. We note that the
choice $\kappa=1/2$ and $Q=\pi/2$ is special since then all the
``small'' Fourier space regions $\Lambda^*_{r,s}$ are identical and
equal to the Brillouin zone of a ``large'' lattice with sites ${\bf
  X}=n_1(1,1)a+n_2(-1,1)a$, $n_j$ integers, which contains eight sites
of the original lattice per elementary cell. However, it is important
to note that our eight flavor model is not local on this ``large''
lattice.

Our fermion flavors $(r,s)$ can be naturally divided into four
different classes with different physical behavior: (i) $(+,0)$ and
$(-,0)$ (antinodal fermions), (ii) $(+,+)$, $(-,+)$, $(+,-)$ and
$(-,-)$ (nodal fermions), (iii) $(-,2)$ (in-fermions), and (iv)
$(+,2)$ (out-fermions). It is interesting to note that this division
of fermions in four classes is also implied by symmetry
considerations: The (discrete) rotation and parity transformations in
\Ref{RP} mix our fermion flavors $(r,s)$ as follows,
\begin{equation}
  \label{RP1}
\begin{split}
  \mathcal{R}:&\; (r,0)\to(-r,0), \quad (r,\pm)\to(\pm r,\mp),
  \quad (r,2)\to (r,2)\\
  \mathcal{P}:&\; (r,0)\to(r,0),\quad (r,\pm)\to(-r,\pm),\quad
  (r,2)\to (r,2)
\end{split}
\end{equation}
for $r=\pm$. Thus the four fermion classes above transform under
different irreducible representations of the group generated by ${\cal
  R}$ and ${\cal P}$.

\subsection{Partial continuum limit I}
\label{sec5.2}
We indent to modify the short-distance details of the eight flavor
model such that it becomes amenable to exact, non-perturbative
computations.  For that we follow a strategy which has been
successfully used in 1D: We assume that there is some underlying Fermi
surface dominating the low energy physics, and we can modify, ignore
or add degrees of freedom far away from this Fermi surface (in the
latter two cases we need to correct the definition of doping, of
course); see Section~\ref{sec3.2} for physical arguments motivating
this strategy.

We expand the dispersion relations of the different fermion flavors
$(r,s)$ in Taylor series in $k_\pm$, and we denote the lowest order
non-trivial terms as $\EE_{r,s}$,
\begin{equation*}
  \epsilon(\vQ_{r,s}+\vk)=\epsilon(\vQ_{r,s}) + \EE_{r,s}(\vk)+\ldots 
\end{equation*}
with the dots indicating higher order terms. We find
\begin{equation}
\label{era} 
\epsilon(\vQ_{r,0})=4t',\quad 
\epsilon(\vQ_{r,\pm})=  -4[t\cos(Q)+t'\cos^2(Q)],
\quad  \epsilon(\vQ_{r,2})= 4(rt-t')
\end{equation}
and
\begin{equation}
\label{Era}
\begin{split}
\EE_{r,0}(\vk) &=  - rc_F k_+k_-- c_F'(k_+^2+k_-^2), \\
\EE_{r,\pm}(\vk)&=  r v_F k_\pm, \\ 
\EE_{r,2}(\vk) &= (-\half rc_F+c_F')(k_+^2+k_-^2)
\end{split} 
\end{equation}
with the constants in \Ref{const}. We see that the nodal fermions
($s=\pm$) have dispersion relations approximately linear and with a
constant Fermi velocity $v_F$, but the dispersion relations of the
antinodal fermions ($s=0$) are quadratic. The in- and out fermions
($s=2$ with $r=-$ and $r=+$, respectively) have energies far away from
the Fermi energy, and we therefore expect that they can be
ignored. This will be done below in Approximation~A3'.

Our first approximation is in the free part of the Hamiltonian:

\begin{itemize}
\item[\textbf{A1:}] \textit{Replace the exact dispersion relations
    $\epsilon(\vQ_{r,s}+\vk)$ in $H_0$ by the truncated Taylor series
    $\epsilon(\vQ_{r,s})+\EE_{r,s}(\vk)$.}
\end{itemize}

This approximation is only essential for the nodal fermions
\cite{dWL1}, and we use it for the other fermion flavors only for
aesthetic consistency.

\begin{table}[b!]
  \begin{center}
    \begin{tabular}{|c||c|c|c|c||c|}
      \hline & & & & &
      \\[-1.0ex] 
      & $r_1,s_1$ & $r_2,s_2$ & $r_3,s_3$& $r_4,s_4$ & Restrictions 
      \\[1.2ex]
      \hline & & & & &
      \\[-1.0ex]
      1. & $r,s$ & $r,s$ & $r',s'$ & $r',s'$ & $(r,s)\neq (r',s')$, $s,s'=0,\pm,2$
      \\
      2. & $r,s$ & $r',s'$ & $r',s'$ & $r,s$ & $(r,s)\neq (r',s')$, $s,s'=0,\pm,2$
      \\
      3. & $r,s$ & $r,s$ & $r,s$ & $r,s$ & $s =0,\pm,2$
      \\
      4. & $r,s$ & $-r,s'$ & $-r,s$ & $r,s'$ & $(s,s')=(\pm,\mp), (0,2), (2,0)$ 
      \\
      5. & $r,s$ & $r,s'$ & $-r,s$ & $-r,s'$ & $(s,s')=(\pm,\mp), (0,2), (2,0)$ 
      \\
      6. & $r,s$ & $-r,s$ & $r,s$ & $-r,s$ & $s=0,2$ 
      \\
      7. & $r,s$ & $r',s'$ & $r,s$ & $-r',s'$ & $s=0,2$, $s'=\pm$
      \\
      8. & $r,s$ & $r',s'$ & $-r,s$ & $r',s'$ & $s=\pm$, $s'=0,2$
      \\
      9. & $r,s$ & $r',s'$ & $r,s$ & $r',s'$ & $s,s'=0,2$
      \\
      10. & $r,s$ & $r',s'$ & $-r',s'$ & $-r,s$ & $(s,s')=(0,2), (2,0)$
      \\
      11. & $r,s$ & $-r,s$ & $r',s'$ & $-r',s'$ & $(s,s')=(0,2), (2,0)$
      \\
      [1.2ex] 
      \hline
    \end{tabular}
  \end{center}  
  \caption{List of all interactions terms potentially contributing to
    the effective model derived in Section~\ref{sec5.2}; $r,r'=\pm$.}
\label{table1}  
\end{table}

Our second approximation is for the interaction part of the
Hamiltonian (again this approximation is only essential for terms
involving nodal fermions).

\begin{itemize}
\item[\textbf{A2:}] \textit{Replace the interaction vertex \Ref{vv8}
    in $\VV$ by
\begin{equation}
\label{vv81}
\begin{split}
  \hat{v}(K_1,K_2,K_3,K_4) = \left(\mbox{$\frac{L}{2\pi}$}\right)^2
  \delta_{\vk_1-\vk_2+\vk_3-\vk_4,\vzero} \, \hat{u}(\vQ_{r_1,s_1}
  -\vQ_{r_2,s_2})
  \\
  \times \sum_{\vn\in\Z^2} \delta_{\vQ_{r_1,s_1}-\vQ_{r_2,s_2}+
    \vQ_{r_3,s_3} -\vQ_{r_4,s_4},2\pi /a\vn} .
\end{split} 
\end{equation} 
}
\end{itemize}

This corresponds to two changes: Firstly, the original interaction
vertex in \Ref{vv8} allows all scattering processes such that
\begin{equation*}
\vQ_{r_1,s_1} - \vQ_{r_2,s_2} + \vQ_{r_3,s_3} -
\vQ_{r_4,s_4} +\vk_1-\vk_2+\vk_3-\vk_4 \in \frac{2\pi}a \Z^2
\end{equation*} 
but we now enforce, in addition, 
\begin{equation}
\label{QQQQ} 
\vQ_{r_1,s_1} - \vQ_{r_2,s_2} + \vQ_{r_3,s_3} -
\vQ_{r_4,s_4} \in \frac{2\pi}a \Z^2. 
\end{equation} 
Secondly, we modify the argument of the weight function $\hat u$ in
\Ref{vv8} by ignoring the dependence on $\vk_1-\vk_2$. Since the
$\vQ_{r_j,s_j}$ are large and the $\vk_j$ (typically) small on the
scale of our short-distance cutoff $\pi/a$, we expect that the
difference between the original- and our approximate interaction does
not change the low energy properties of the model. 

We therefore only keep the interaction terms between the fermion
flavors for which the condition in \Ref{QQQQ} is fulfilled. For
$Q=\pi/2$ we find 512 such terms, but for $Q\neq\pi/2$ there are only
the 196 terms listed in table~\ref{table1}.  Below we discuss the
corresponding interactions in more detail.

Terms 1 in table~\ref{table1} are Hartree-like, i.e.\ they describe
scattering processes $\vQ_{r,s}\to\vQ_{r,s}$,
$\vQ_{r',s'}\to\vQ_{r',s'}$ which preserve fermion flavors. They are
given by
\begin{equation*}
\VV_1 = \sum'_{r,s, r',s'} \sum_{\vk_j} (\tPiL)^6 \hat{u}(\vzero)
\delta_{\vk_1-\vk_2+\vk_3-\vk_4,\vzero}\, \hat\psi^\dag_{r,s}(\vk_1)
\hat\psi^{\phantom\dag}_{r,s}(\vk_2) \hat\psi^\dag_{r',s'}(\vk_3)
\hat\psi^{\phantom\dag}_{r',s'}(\vk_4)
\end{equation*}
where the prime in the sum indicates that the diagonal terms
$(r,s)=(r',s')$ are excluded (the latter are terms 3 in
table~\ref{table1} and are treated separately). It is useful to
introduce the operators
\begin{equation}
\label{hrho}
\hat\rho_{r,s}(\vp) \define \sum_{\vk_1,\vk_2\in\Lambda^*_{r,s}} (\tPiL)^2 
\hat\psi^\dag_{r,s}(\vk_1)\hat\psi^{\phantom\dag}_{r,s}(\vk_2)
\delta_{\vk_1+\vp,\vk_2} 
\end{equation}
which allow to write the Hartree-like interactions in the following
simpler form,
\begin{equation}
\label{V1}
\VV_1 = \sum'_{r,s, r',s'}\sum_{\vp}  (\mbox{$\frac{a}{L}$})^2
V \hat\rho_{r,s}(-\vp)\hat\rho_{r',s'}(\vp)
\end{equation}
where we used $(2\pi)^2\hat{u}(\vzero)=a^2 V$. The operators in
\Ref{hrho} can be naturally interpreted as Fourier transformed fermion
densities.

Terms 2 are Fock-like, i.e.\ the corresponding scattering processes
$\vQ_{r,s}\to\vQ_{r',s'}$, $\vQ_{r',s'}\to\vQ_{r,s}$ exchange fermion
flavors. They are given by
\begin{equation*}
\VV_2 = \sum'_{r,s, r',s'} \sum_{\vk_j} (\tPiL)^6
\hat{u}(\vQ_{r,s}-\vQ_{r',s'})
\delta_{\vk_1-\vk_2+\vk_3-\vk_4,\vzero}\, \hat\psi^\dag_{r,s}(\vk_1)
\hat\psi^{\phantom\dag}_{r',s'}(\vk_2) \hat\psi^\dag_{r',s'}(\vk_3)
\hat\psi^{\phantom\dag}_{r,s}(\vk_4).
\end{equation*}
Renaming $\vk_2,\vk_4$ to $\vk_4,\vk_2$ and using the anticommutator
relations of the fermion field operators we can also write $\VV_2$ in
terms of fermion densities,
\begin{equation}
\label{V2}
\VV_2 = -\sum'_{r,s, r',s'} v_{r,s,r',s'} 
\sum_{\vp} (\mbox{$\frac{a}{L}$})^2 
\hat\rho_{r,s}(-\vp)\hat\rho_{r',s'}(\vp) 
+ \sum'_{r,s, r',s'} v_{r,s,r',s'} 
N_{r,s} f_{r',s'}
\end{equation} 
with $N_{r,s}=\hat\rho_{r,s}(\vzero)$ equal to the number operators in
\Ref{Nrs} and the constants
\begin{equation*}
  v_{r,s,r',s'} \define (2\pi)^2  
  \hat{u}(\vQ_{r,s}-\vQ_{r',s'})/a^2=v_{r',s',r,s}
\end{equation*} 
and $f_{r,s}\define \sum_{\vk\in\Lambda^*_{r,s}}(a/L)^2$. By simple
computations we find
\begin{equation}
\label{urara}
\begin{split} 
v_{+,0,-,0} &= v_{+,2,-,2}= -V,\quad v_{+,\pm,-,\pm} = 
V\cos(2Q) \\ 
v_{r,0,r',\pm} &= v_{r,0,r',2} = 0,\quad v_{r,\pm ,r',\mp} = 
\mbox{$\frac12$}V[1+\cos(2Q)] \\
v_{r,\pm,-,2}&=V\cos(Q),\quad v_{r,\pm,+,2}=-V\cos(Q),
\quad v_{r,s,r,s}=V 
\end{split} 
\end{equation}
and 
\begin{equation}
\label{fra}
f_{r,0} = \half\kappa^2,\quad 
f_{r,\pm}=\half\kappa(1-\kappa),\quad 
f_{r,2}=\half(1-\kappa)^2
\end{equation} 
for all $r,r'=\pm$.  Note that $f_{r,s}$ equals the ratio of the area
of $\Lambda^*_{r,s}$ to the area of $\BZ$, and $\sum_{r,s}f_{r,s}=1$.

We excluded the diagonal Hartree terms where $(r,s)=(r',s')$ above
since, by simple computations similar to the ones described above,
they can be simplified to
\begin{equation}
\label{V3} 
\VV_3 = \sum_{r,s}\sum_{\vk,\vk'} (\tPiL)^4 \hat{u}(\vzero) 
\hat\psi^\dag_{r,s}(\vk)\hat\psi_{r,s}^{\phantom\dag}(\vk)
= \sum_{r,s}V N_{r,s} f_{r,s} 
\end{equation} 
and thus do not contribute to the interactions; see
Appendix~\ref{appA} for details.

Terms 4 and 5 in table~\ref{table1} are mixed, i.e.\ they are
Hartree-like in one and Fock-like in the other of the components of
the momenta. By straightforward computations one finds that they
exactly add up to zero; see Appendix~\ref{appA} for details.

Terms 6 are back-scattering terms where the condition in \Ref{QQQQ}
holds true for a non-zero integer vector $\vn$. A simple computations
shows that these terms are identical to zero; see Appendix~\ref{appA}.

Terms 7--9 are BCS-like, i.e.\ $\vQ_{r_1,s_1}+\vQ_{r_3,s_3}\in
2\pi\Z^2/a$ and $\vQ_{r_2,s_2}+\vQ_{r_4,s_4}\in 2\pi\Z^2/a$ (note that
$\vQ_{r,s}+\vQ_{r,s}\in 2\pi\Z^2/a$ for $s=0,2$ and
$\vQ_{r,s}+\vQ_{-r,s}=\vzero$ for $s=\pm$). Straightforward
computations show that all these terms are identically zero; see
Appendix~\ref{appA}.  There are also BCS-like terms
$(r_1,s_1)=(r,\pm)$, $(r_2,s_2)=(-r,\pm)$, $(r_3,s_3)=(-r,\pm)$,
$(r_4,s_4)=(r,\pm)$ which give non-zero contributions, but they are
already included in the Fock terms 2 above.

Terms 10 and 11 are back-scattering terms. One can show that the
former vanish, and the latter are equal to
\begin{equation}
\label{V11}
\VV_{11} = - V \sum_{\vk_j} \sum_{r,r'} \left(\tPiL\right)^{4} 
(\mbox{$\frac{a}{L}$})^2 
\hat\psi^\dag_{r,0}(\vk_1) \hat\psi^{\phantom\dag}_{-r,0}(\vk_2)
\hat\psi^\dag_{r',2}(\vk_3)
\hat\psi^{\phantom\dag}_{-r',2}(\vk_4); 
\end{equation}
see Appendix~\ref{appA}.

For $Q=\pi/2$ the equation in \Ref{QQQQ} has many more solutions. This
leads to additional interactions, including, for example,
\begin{equation*}
  \VV_{\mathrm{BS}}= V\sum_{\vk_j} \sum_{r,r'} \sum_{s=\pm} \left(\tPiL\right)^{4} 
  (\mbox{$\frac{a}{L}$})^2 
  \hat\psi^\dag_{r,0}(\vk_1) \hat\psi^{\phantom\dag}_{-r,0}(\vk_2)
  \hat\psi^\dag_{r',s}(\vk_3)
  \hat\psi^{\phantom\dag}_{-r',s}(\vk_4).  
\end{equation*} 
Nodal back-scattering terms like this, involving nodal fermion
operators but which cannot be written in terms of nodal densities
$\hat\rho_{r,\pm}(\vp)$, cannot be bosonized in a simple manner. In
the following we therefore assume that $Q\neq \pi/2$.

To summarize, we found that, using the Approximations~A1 and A2 and
assuming $Q\neq \pi/2$, the free- and interaction parts of the 2D
\ttpV model Hamiltonian are equal to
\begin{equation}
  H_0 = \sum_{r,s} \sum_{\vk} (\tPiL)^2 {[}\epsilon(\vQ_{r,s}) + 
  \EE_{r,s}(\vk) -\mu{]} \hat\psi^\dag_{r,s}(\vk)
  \hat\psi^{\phantom\dag}_{r,s}(\vk) 
\end{equation}
and $\VV = \VV_1+\VV_2+\VV_3+\VV_{11}$, i.e.
\begin{equation}
\begin{split}
  \VV = \sum'_{r,s,r',s'} (V-v_{r,s,r',s'}) \sum_{\vp}
  (\mbox{$\frac{a}{L}$})^2 \hat\rho_{r,s}(-\vp) \hat\rho_{r',s'}(\vp)
  +
  \sum_{r,s,r',s'} v_{r,s,r',s'} N_{r,s}f_{r',s'} +\VV_{11} 
\end{split}
\end{equation} 
with the dispersion relations in \Ref{Era} and the constants in
\Ref{const}, \Ref{era}, \Ref{urara} and \Ref{fra}. In the following we
denote the full Hamiltonian $H_0+\VV$ as $H+\VV_{11}$ (since
$\VV_{11}$ is not affected by the computations in the next section and
will be dropped in Approximation~A3').

 \bigskip

\noindent \textbf{Remark:} The fact that many of the interaction terms
listed in table~\ref{table1} add up to zero is a consequence of the
Pauli exclusion principle and holds true since we use spinless
fermions.

\subsection{Normal ordering}
\label{sec5.3}
Up to now we assumed a reference state $|0\rangle$ in the fermion Fock
space which is annihilated by all fermion operators
$\hat\psi(\vk)$. Before we can partly remove the UV cutoff below
(Approximation~A4) it is important to introduce another reference
state $\Omega$ in which all momentum states $\vk$ inside some Fermi
surface are filled.  This state can be defined as follows,
\begin{equation*}
\Omega \define \prod_{\vk} \hat\psi^\dag(\vk)|0\rangle
\end{equation*}
where the product is over the set of all filled momenta $\vk$
specified below.  After Approximation~A4 below one can only measure
physical quantities with respect to this state $\Omega$, i.e., only
expectation values of normal ordered operators are finite. We
therefore have to determine the effect of normal ordering for various
operators of interest to us.

To be more specific, we assume that nodal fermion states ($s=\pm$) are
filled up to the Fermi surface arcs described in Section~\ref{sec3.2}
(Hypothesis~H2) and indicated by dashed lines in figure~\ref{FIG1}.
This is the only assumption we need to make about the Fermi surface
since the partial continuum that we perform below only affects the
nodal fermions.  However, it is also convenient to assume that the
antinodal fermions ($s=0$ and $r=\pm$) have some filling $\nu_a/2$
which is the same for $r=+$ and $r=-$.  We expect that there is an
interesting parameter regime with $\nu_a=1/2$ \cite{EL0}, but this
remains to be shown \cite{dWL1}. It is also convenient to assume that
the in- and out fermions ($s=2$ with $r=-$ and $r=+$, respectively)
are totally filled and totally empty, respectively.  Thus the sets of
filled states for the different fermion flavors $r,s$ are as follows,
\begin{equation*}
  S_{r,\pm} = \left\{ \vk\in\Lambda^*_{r,\pm}:\;  rk_\pm <0\right\},\quad 
    S_{-,2}=\Lambda^*_{-,2},\quad S_{+,2} = \emptyset 
  \end{equation*}
and $S_{r,0}$ such that the r.h.s.\ of \Ref{nua} is the same for $r=+$
and $r=-$ (as will be seen, we need not be more specific about
$S_{r,0}$).  This allows us to fully characterize the new reference
state as follows,
\begin{equation*}
  \hat\psi^\dag_{r,s}(\vk)\Omega =0\quad \forall \vk\in S_{r,s}, \quad 
  \hat\psi^{\phantom \dag}_{r,s}(\vk)\Omega =0 \quad \forall 
\vk\notin S_{r,s}
\end{equation*}
leading to \Ref{DiracSea} after Approximation~A4 below.  We compute
the contributions $\nu_{r,s}\define (a/L)^2 \langle\Omega,
N_{r,s}\Omega\rangle$ of the different fermion flavors to the filling
factor in this new reference state and find
\begin{equation}
\label{nura}
  \nu_{r,0} = \half\nu_a\kappa^2,\quad 
  \nu_{r,\pm} =  \quart(1-\kappa)\left(\QQPi-1+\kappa\right),\quad  
\nu_{-,2}=\half(1-\kappa)^2, \quad \nu_{+,2}= 0  
\end{equation} 
with $\nu_a$ in \Ref{nua}, and for the total filling $\sum_{r,s}
\nu_{r,s}$ we obtain the result in \Ref{nutot1}.

We normal order various operators quadratic in the fermion fields as
usual; see \Ref{NO}.  In particular, the normal ordered fermion number
operators are $:\!N_{r,s}\!:\, = N_{r,s} - (L/a)^2 \nu_{r,s}$. Thus
the filling is related to the expectation values of these normal
ordered particle number operators as follows,
\begin{equation}
\label{nutot} 
\nu = \half + (1-\kappa)\left(\QQPi-1\right)  + \kappa^2(\nu_a-\half)
+ \mbox{$(\frac{a}{L})^2$}\sum_{r,s} 
\langle\, :\!N_{r,s}\!:\, \rangle. 
\end{equation} 

To compute the Hamiltonian in its normal ordered form we have to write
it in terms of normal ordered fermion densities
$\hat{J}_{r,s}(\vp)\define \; :\!\hat\rho_{r,s}(\vp)\!:$. Since normal
ordering of these densities is irrelevant unless $\vp=0$ and
$\hat\rho_{r,s}(\vzero)=N_{r,s}$ we obtain
\begin{equation}
\hat{J}_{r,s}(\vp) = \hat\rho_{r,s}(\vp) -
\delta_{\vp,\vzero}\,
\left(\mbox{$\frac{L}{a}$}\right)^2\nu_{r,s} .
\label{hatJrs1}
\end{equation} 
Inserting this into the Hamiltonian $H$ obtained at the end of the
previous section we get
\begin{equation}
  \label{normalH}
\begin{split}
  H =& \sum_{r,s} \sum_{\vk\in\Lambda^*_{r,s}} (\tPiL)^2
  {[}\EE_{r,s}(\vk)-\mu_{r,s}{]} :\!
  \hat\psi^\dag_{r,s}(\vk)\hat\psi^{\phantom\dag}_{r,s}(\vk)\!:\,
  \\
  &+ \sum'_{r,s,r',s'} (V-v_{r,s,r',s'}) \sum_{\vp\in\tilde\Lambda^*}
  (\mbox{$\frac{a}{L}$})^2 \hat{J}_{r,s}(-\vp)\hat{J}_{r',s'}(\vp) +
  \cE_0
\end{split} 
\end{equation} 
with the dispersion relations in \Ref{Era}, the parameters
\begin{equation}
\label{murs} 
\mu_{r,s}=\mu- \epsilon(\vQ_{r,s}) - 2V\sum_{r',s'}\nu_{r',s'} 
-\sum_{r',s'}v_{r,s,r',s'}(f_{r',s'}-2\nu_{r',s'}) ,  
\end{equation}
and the constants in \Ref{const}, \Ref{era}, \Ref{urara}, \Ref{fra}
and\footnote{The $r,s$ summations in the following formulas are
  written explicitly since they are not changed in Approximation~A3'
  below.}
\begin{equation}
  \cE_0 = \sum_{r=\pm}\sum_{s=0,\pm,2}\sum_{\vk\in\Lambda^*_{r,s}} (\tPiL)^2 
  {[}\epsilon(\vQ_{r,s}) + \EE_{r,s}(\vk){]} 
  \langle\Omega, 
  \hat\psi^\dag_{r,s}(\vk)\hat\psi^{\phantom\dag}_{r,s}(\vk)\Omega\rangle 
  + \cE_{int} 
\end{equation} 
with
\begin{equation}
  \cE_{int} = L^2 \left( V\nu^2 -\mu\nu + 
    \sum_{r,r'=\pm}\sum_{s,s'=0,\pm,2}
    v_{r,s,r',s'}\nu_{r,s}(f_{r',s'}-\nu_{r',s'} ) \right) 
\end{equation} 
and $\nu$ in \Ref{nutot1}. The constants $\mu_{r,s}$ correspond to
chemical potentials which are different for the different fermion
flavors.

The condition that the points $\vQ_{r,\pm}$ are on the Fermi surface
is equivalent to $\mu_{r,\pm}=0$, and this fixes the chemical
potential $\mu$ and $\mu_{r,s}$ for $r=\pm$ and $s=0,2$.  By
straightforward computation we find $\mu_{\pm,0}=\mu_0$ in \Ref{mua}
and
\begin{equation}
\label{mupm2}  
\mu_{\pm,2}  =  
\mp[4t+V(1-\kappa)(1-\kappa+2(\QQPi-1)\cos(Q))][1\pm\cos(Q)] 
+4t'\sin^2(Q) .
\end{equation}

The particle-hole transformation mentioned in Section~\ref{sec4.1}
provides a useful check of our computations above: under this
transformation, $\hat\psi^{\phantom\dag}_{r,s}(\vk) \leftrightarrow
\hat\psi^\dag_{-r,s}(\vk)$, and this is equivalent to the change of
parameters in \Ref{ParticleHole} and
\begin{equation*}
(Q,\nu_{r,s}) \to (\pi-Q,f_{-r,s}-\nu_{-r,s}). 
\end{equation*}
Our approximate model above is invariant under this transformation
since, in addition, $\mu_{r,s}\to -\mu_{-r,s}$ and $\cE_0\to\cE_0$.

\bigskip

\noindent \textbf{Remark:} A careful reader might wonder if one could
also work with the normal ordered form of the interaction in \Ref{V0},
i.e.
\begin{equation*}
  \VV=\sum_{\vx,\vy} a^4 u(\vx-\vy) \psi^\dag(\vx)\psi^\dag(\vy) 
  \psi(\vy)\psi(\vx)
\end{equation*}
(this would make the arguments in Section~\ref{sec5.2} somewhat
simpler). The answer is ``no'' since Approximation~A2 would spoil
particle-hole transformation symmetry then.

\subsection{Partial continuum limit II}
\label{sec5.4}
We assume that the parameters are such that
\begin{equation}
\label{condition} 
\mu_{-,2} \ll 0,\quad \mu_{+,2} \gg 0 
\end{equation} 
on an energy scale of order $t/5$ or so.  This gives some restrictions
on the allowed parameter values that are important to check. If these
conditions are fulfilled the in- and out fermions ($s=2$ with $r=-$
and $r=+$, respectively) have energies sufficiently far below and far
above the Fermi energy, respectively, that the following approximation
is expected to be appropriate.\footnote{In the simplified treatment in
  \cite{EL0} the following approximation was included in A4 below, and
  for this reason we label it as A3' rather than A3.}

\begin{itemize}
\item[\textbf{A3':}] \textit{Drop all terms in the total Hamiltonian
    involving in- or out fermions.}
\end{itemize}

In particular we drop the interaction term $\VV_{11}$. Thus the total
Hamiltonian is as in \Ref{normalH}, but from now on sums over $r,s$
are restricted to $r=\pm$ and $s=0,\pm$ unless stated otherwise.  We
thus obtain a model involving only nodal and antinodal fermions.

We recall that $\hat{J}_{r,s}(\vp)$ is proportional to the sum of all
terms $:\!\hat\psi^\dag_{r,s}(\vk_1)\hat\psi_{r,s}(\vk_2)\!:$ with
momenta $\vk_j\in\Lambda^*_{r,s}$ such that $\vp=\vk_2-\vk_1$. Since
the $\vk_j$ belong to finite sets the number of these terms gets
smaller as $|p_\pm|$ is increased, and it vanishes for $\vp$ outside
some finite set. In particular, $\hat{J}_{r,s=\pm}(\vp)$ is non-zero
only if $|p_s|\leq \sqrt{2}\kappa\pi/a$ and $|p_{-s}|\leq 
\sqrt{2}(1-\kappa)\pi/a$. In our Approximations~A4 and A5 below we add
terms to $\hat{J}_{r,s=\pm}(\vp)$. The following approximation is to
partly compensate for this.

\begin{itemize}
\item[\textbf{A3:}] \textit{Replace $\hat{J}_{r,s=\pm}(\vp)$ in
    \Ref{normalH} by $\chi_s(\vp)\hat{J}_{r,s}(\vp)$ with
    $\chi_s(\vp)$ the cutoff functions in \Ref{chis}.}
\end{itemize}

In the following approximation we partly remove the short-distance
cutoff $a$ for the nodal fermions by dropping the restriction on $k_s$
in $\Lambda^*_{r,s=\pm}$:

\begin{itemize}
\item[\textbf{A4:}] \textit{Replace the nodal Fourier spaces
    $\Lambda^*_{r,s=\pm}$ in \Ref{Lambdars} by $\Lambda^*_{s=\pm}$ in
    \Ref{Lambdas}.}
\end{itemize}

The final approximation adds terms to the nodal interactions and is
necessary to obtain a model that can be bosonized in a simple manner.

\begin{itemize}
\item[\textbf{A5:}] \textit{Replace the nodal fermion densities thus
    obtained
    \begin{equation}
      \hat{J}_{r,s=\pm}(\vp)  = \sum_{\vk\in\Lambda^*_s}(\tPiL)^2
      :\!
      \hat\psi^\dag_{r,s}(\vk-\vp)\hat\psi\pdag_{r,s}(\vk)\!: 
  \label{hatJrs2}
\end{equation}
by the ones in \Ref{hatJrs}, i.e., insert the integer sum.}
\end{itemize}

We thus obtain the Hamiltonian of the 2D Luttinger model as defined in
Section~\ref{sec4.2} (To simplify our notation we introduced the
constants in \Ref{gj} which we obtained by computing
$2a^2(V-v_{r,\pm,-r,\pm})=g_1$ etc.)  Approximation~A5 amounts to
adding umklapp terms in the $k_{-s}$-direction. Since these umklapp
terms vanish in the full continuum limit $a\to 0$ (at fixed parameters
$v_F$ etc.) we expect them to not (much) affect the low energy
properties of the model.

\bigskip 

\noindent \textbf{Remark:} In Approximation~A3 it is important to use
some cutoff function $\chi_s(\vp)$ to get a model that is
mathematically well-defined. However, one has some freedom there: one
could equally well replace the cutoff function in \Ref{chis} by
\begin{equation*}
  \chi_{s=\pm}(\vp)=\begin{cases} 1 & \mbox{ if }\; 
    |p_s|\leq \frac{b_1\kappa\pi}{\sqrt{2}a}
    \; \mbox{ and }\;  |p_{-s}|\leq \frac{b_2(1-\kappa)\pi}{\sqrt{2}a}\\
    0 & \mbox{ otherwise} \end{cases} 
\end{equation*}
with additional free parameters $b_j>0$. According to Hypothesis~H1 in
Section~\ref{sec3.2} the precise value of these parameters $b_j$ is
not important. Our discussion above suggests $b_j\leq 2$. Moreover, we
find it natural to choose $b_1=b_2$, and the fact that
$\hat{J}_{s=\pm,r}(\vp)$ is defined for $|p_{-s}|\leq
(1-\kappa)\pi/(\sqrt{2}a)$ motivates our choice $b_2=1$. We note that
the choice $b_1\kappa=b_2(1-\kappa)$ would be more convenient since it
leads to simpler formulas.

\subsection{Position space regularizations}
\label{sec5.5} 
The fermion operators used in Section~\ref{sec2} are related to the
ones in the present section by Fourier transformation
\begin{equation}
\label{FT}
\psi_{r,s}(\vx)= \frac1{2\pi} \sum_{\vk\in\Lambda^*_{s}} 
(\tPiL)^2 \hat\psi_{r,s}(\vk)\ee^{\ii\vk\cdot\vx}.     
\end{equation}
The position vectors $\vx$ therefore live on different spaces for
different values of $s$: $\Lambda^*_{s=\pm}$ is the Fourier space
corresponding to spatial positions $\vx$ where $x_s$ is continuous but
$x_{-s}$ in a 1D lattice $\Lambda_{\mathrm{1D}}$ with lattice constant
in \Ref{tildea} and $L/\tilde{a}$ sites,\footnote{Note that
  \Ref{Qquant} ensures that $L/(2\tilde{a})\in(\N+\half)$.}
\begin{equation*}
  \Lambda_{\mathrm{1D}}\define\left\{ x=n\tilde{a} 
    :\; n\in\Z,\quad |n| < L/(2\tilde{a}) \right\} .
\end{equation*}
The precise meaning of $ H_0^\pm \define \int d^2 x \sum_r rv_F
:\!\psi^\dag_{r,\pm}(\vx)(-\ii \partial_\pm)\psi_{r,\pm}^{\phantom\dag}(\vx)\!:
$ is therefore
\begin{equation}
\label{H0pm} 
H_0^\pm = \int_{-L/2}^{L/2} dx_\pm 
\sum_{x_{\mp}\in\Lambda_{\mathrm{1D}}} \tilde{a}
\sum_{r=\pm} rv_F
:\! 
\psi^\dag_{r,\pm}(\vx)(-\ii \partial_\pm)\psi_{r,\pm}^{\phantom\dag}(\vx)\!:   
\end{equation}
(Riemann sum in $x_\mp$) and the precise meaning of
$\{\psi^{\phantom\dag}_{r,\pm}(\vx), \psi^\dag_{r,\pm}(\vy) \} =
\delta^2(\vx-\vy)$ is
\begin{equation}
\label{deltareg}
  \{  \psi^{\phantom\dag}_{r,\pm}(\vx), \psi^\dag_{r,\pm}(\vy)\} 
  = \frac1{\tilde{a}}\delta_{x_\mp,y_\mp} \delta(x_\pm-y_\pm).  
\end{equation}
The reason for adding the umklapp terms in Approximation~A5 is that
\Ref{FT} and \Ref{hatJrs} for $s=\pm$ imply that
\begin{equation}
  J_{r,s=\pm}(\vx) = \sum_{\vp\in\C_s} 
  \fLL \hat{J}_{r,s}(\vp)\ee^{\ii\vp\cdot\vx} = 
  \, :\!\psi^\dag_{r,s}(\vx)\psi\pdag_{r,s}(\vx)\!:, 
\end{equation} 
i.e.\ the nodal fermion densities are local. Note that there is
another UV cutoff suppressed in the formal nodal interaction in
\Ref{Hn1}, namely
\begin{equation*}
  \VV_n = 2\tint d^2x \tint d^2y \left( g_1 \sum_{s=\pm}
    \tilde\delta_{s,s}(\vx-\vy) J_{+,s}(\vx)J_{-,s}(\vy) +
    g_2\sum_{r,r'=\pm} \tilde\delta_{+,-}(\vx-\vy)
    J_{r,+}(\vx)J_{r',-}(\vy) \right)
\end{equation*} 
with $\tint d^2x$ short for
$\int_{-L/2}^{L/2}dx_s\sum_{x_{-s}\in\Lambda_{\mathrm{1D}}}\tilde{a}$
(for appropriate $s$) and
\begin{equation*}
  \tilde\delta_{s,s'}(\vx-\vy) = \sum_{\vp\in\tilde\Lambda^*} \fLL
  \chi_s(\vp)\chi_{s'}(\vp')\ee^{-\ii\vp\cdot(\vx-\vy)}
\end{equation*}
approximate Dirac delta functions. The formulas above allow to make
the nodal Hamiltonian in \Ref{Hn1} precise as $H\pdag_n=H_n^+ + H_n^-
+ H^\prime_n$. The precise meaning of the antinodal and mixed parts in
\Ref{Ha1} and \Ref{Hna1} can be given along the same lines but is not
needed in the following.

\section{Bosonization and partial exact solution}
\label{sec6} 
In this section we show that the 2D Luttinger model derived in the
last section is amenable to analytical, non-perturbative computations.

\subsection{Bosonization}
\label{sec6.1}
The Hamiltonians in \Ref{H0pm} can be interpreted as a model for 1D
fermions with $x_\pm$ a continuum spatial variable and $x_\mp$ a
discrete flavor index. It is therefore possible bosonize the
Hamiltonian using the same mathematical results that allow to bosonize
the 1D Luttinger model, as explained in more detail in
Appendix~\ref{appB}. We collect these results, adapted to our needs,
in the following.

\noindent \textbf{Proposition:}
\label{prop1} {\it(a) The nodal density operators are well-defined
  operators obeying the commutator relations
\begin{equation} 
  \label{Propa1}
  [\hat{J}_{r,\pm}(\vp), \hat{J}_{r',\pm}(\vp')]= r\delta_{r,r'}
  \frac{2\pi p_s}{\tilde{a}}\delta_{\vp,-\vp'}(\mbox{$\frac{L}{2\pi}$})^2
\end{equation}
and $[\hat{J}_{r,\pm}(\vp), \hat{J}_{r',\mp}(\vp')]=0$. Moreover,
$\hat{J}_{r,\pm}(\vp)^\dag=\hat{J}_{r,\pm}(-\vp)$ and
\begin{equation}
\label{Propa2}
\hat{J}_{r,\pm}(\vp)\Omega=0\quad \forall \vp\, \mbox{ such that }\, 
rp_{\pm}\geq 0. 
\end{equation}

\noindent (b) The following identity holds true
\begin{equation}
\label{Propb}
  \sum_{\vk\in\Lambda^*_{\pm}} (\tPiL)^2 rk_\pm :\!
  \hat\psi^\dag_{r,\pm}(\vk)\hat\psi\pdag_{r,\pm}(\vk)\!:\; 
  = \tilde{a}\pi \sum_{\vp\in\C_\pm} \fLL :\! 
  \hat{J}_{r,\pm}(-\vp) \hat{J}_{r,\pm}(\vp)\!: 
\end{equation} 
with $\C_\pm$ in \Ref{Lams}.}

\noindent (Proof outlined in Appendix~\ref{appB} with details in
\cite{dWL2}.)

Part (a) of this proposition implies that the operators
\begin{equation*}
  b_{s=\pm}(\pm \vp) \define\mp\frac{\ii}{L}\sqrt{\frac{2\pi\tilde{a}}{p_s}}
  \hat{J}_{\pm,s}(\pm\vp)\quad \forall \vp \; \mbox{ such that }\;
  p_s>0
\end{equation*}
are standard 2D boson operators, i.e.\ they obey the usual commutator
relations
\begin{equation*}
  [b\pdag_{s}(\vp),b_{s'}^\dag(\vp')]=\delta\pdag_{s,s'}
  \delta^{\phantom\dag}_{\vp,\vp'}
\end{equation*} 
etc.\ and $b_{\pm}(\vp)\Omega = 0$. Part (b) implies that the nodal
Hamiltonian $H_n$ in \Ref{Hn} is identical with a non-interacting
boson Hamiltonian, and \Ref{Hna} shows that the antinodal densities
are coupled linearly to these bosons.

We note that the boson operators $b^{(\dag)}_{s=\pm}(\vp)$ above are
only defined if $p_s\neq 0$, and $H_n$ includes also terms that cannot
be expressed in terms of these boson operators due to the zero modes
$\hat{J}_{r,s=\pm}(\vp)|_{p_s =0}$ commuting with all boson
operators. However, $H_n$ scales like $L^2$ for large systems sizes
$L$, whereas the sum of all these zero mode terms scales at most like
$L$. This suggests that, for large systems, these zero mode terms can
be ignored. To simplify our discussion below we therefore ignore zero
mode terms, i.e.\ $H_n$ in the following differs from $H_n$ in
\Ref{Hn} in that all terms with zero modes are implicitly excluded. A
complete solution of the nodal model, including a detailed treatment
of the zero modes, is given elsewhere \cite{dWL2}.

It is important to note that our scaling is such that the doping of
the nodal fermions relative to $\Omega$ is zero in the limit
$L/a\to\infty$; see \Ref{Nn} in Appendix~\ref{appB}. Moreover, the
nodal filling $\nu_a$ is to be determined such that $(a/L)^2:\!
N_{\pm,0}\!:\; =0$.  Thus the doping constraint in \Ref{nutot} can be
simplified to the one in \Ref{nutot1}.

\subsection{Diagonalization of the nodal model} 
\label{sec6.2}
A convenient way to write the bosonized nodal Hamiltonian is in terms
of the operators
\begin{equation}
\label{PhiPi}
\begin{split}
  \hat\Phi_s(\vp)&= \frac1{\ii p_s}
  \sqrt{\frac{\tilde{a}}{4\pi}}[\hat{J}_{+,s}(\vp) +
  \hat{J}_{-,s}(\vp)] \\
  \hat\Pi_s(\vp)&= \sqrt{\frac{\tilde{a}}{4\pi}}[-\hat{J}_{+,s}(\vp) +
  \hat{J}_{-,s}(\vp)]
\end{split}  
\quad (p_s\neq 0)
\end{equation}
which are standard 2D boson operators, i.e.\ they obey the usual
canonical commutator relations
\begin{equation*}
  [\hat\Pi\pdag_s(\vp),\hat\Phi^\dag_{s'}(\vp)]=
-\ii \delta_{s,s'}\delta_{\vp,\vp'}
  (\mbox{$\frac{L}{2\pi}$})^2
\end{equation*} 
etc., with $\Phi_s^\dag(\vp)=\hat\Phi\pdag_s(-\vp)$ and
$\hat\Pi^\dag_s(\vp)=\hat\Pi_s\pdag(-\vp)$. We refer to the
quasi-particles corresponding to the fields in \Ref{PhiPi} as {\em
  nodal bosons}.

Inserting this and \Ref{Propb} in \Ref{Hn} and \Ref{Hna} we obtain
\begin{equation}
\begin{split}
\label{Hn2}
H_n = \frac{v_F}{2}\Bigl( \sum_{s=\pm} \sum_{\vp\in\C_s}(\tPiL)^2 :\!
\bigl( [1-\gamma\chi_s(\vp) ]\hat\Pi^\dag_s(\vp)\hat\Pi_s(\vp)+
[1+\gamma\chi_s(\vp)]p_s^2
\hat\Phi^\dag_s(\vp)\hat\Phi\pdag_s(\vp)\bigr)\!: \\ +
\sum_{\vp\in\C_+\cap \C_-} (\tPiL)^2 \gamma\chi_+(\vp)\chi_-(\vp)
p_+p_-[\hat\Phi^\dag_+(\vp)\hat\Phi\pdag_-(\vp)+
\hat\Phi^\dag_-(\vp)\hat\Phi\pdag_+(\vp)] \Bigr)
\end{split} 
\end{equation} 
and 
\begin{equation}
\label{Hna2}
H_{na} = \frac{g_4}{\sqrt{\pi\tilde{a}}}
\sum_{\vp\in\C_s}\fLL 
\sum_{r=\pm}\hat{J}_{r,0}(-\vp)\sum_{s=\pm} 
 \chi_s(\vp)2\pi\ii p_s \hat\Phi_s(\vp)
\end{equation}
with $\gamma\define g_1/(v_F\tilde{a}\pi)=2g_2/(v_F\tilde{a}\pi)$.
Inserting \Ref{const} and \Ref{tildea} we obtain \Ref{gamma}.

It is straightforward to diagonalize the nodal Hamiltonian in
\Ref{Hn2} by a boson Bogoliubov transformation \cite{dWL2}. We obtain
\begin{equation}
\label{Hndiag}
H_n = \cE_n + \sum_{s=\pm}\sum_{\vp\in\C_s} 
\omega\pdag_s(\vp)\tilde b_s^\dag(\vp)\tilde b_s\pdag(\vp) 
\end{equation}
with
\begin{equation}
\label{Es} 
\omega_{s=\pm}(\vp) = \begin{cases} 
    v_F\sqrt{\frac{1-\gamma^2}2}\sqrt{\left( 
        |\vp|^2 + s \sqrt{|\vp|^4 - A(2p_+p_-)^2 }\right)} & \mbox{if }\; 
    |p_\pm|\leq \pi\kappa_{\min}/(\sqrt{2}a) \\
    v_F|p_s| & \mbox{if }\; |p_s|> \pi\kappa/(\sqrt{2}a)\\
    & \\[-1.5ex] 
    v_F\sqrt{1-\gamma^2}|p_s| & \mbox{otherwise} 
\end{cases} 
\end{equation}
the boson dispersion relation, where we use the shorthand notation
\begin{equation}
  A\define 
  1-\frac{\gamma^2}{(1+\gamma)^2},\quad \kappa_{\min}
  \define \min(\kappa,1-\kappa). 
\end{equation} 
The constant $\cE_n$ is the nodal ground state energy, and we obtain
\begin{equation}
\label{cEn} 
\cE_n = \frac12\sum_{s=\pm} 
\sum_{\vp\in\C_s} [\omega_s(\vp)-v_F|p_s|]
\end{equation}
which is finite since $\omega_s(\vp)-v_F|p_s|$ is non-zero only for a
finite number of terms. The $\tilde b^{(\dag)}_s(\vp)$ are Bogoliubov
transformed boson operators and can be computed explicitly
\cite{dWL2}. From \Ref{Hndiag} is is easy to construct a complete set
of exact eigenstates of $H_n$. This solution shows that $H_n$ is
well-defined if and only if $\gamma<1$, as mentioned earlier.

Note that only a finite number of nodal boson modes are coupled to the
antinodal fermions. Moreover, there are only a finite number of
antinodal fermion degrees of freedom. Thus the 2D Luttinger
Hamiltonian is well-defined if and only if $H_n$ is.

\subsection{Effective antinodal model} 
\label{sec6.3}
In the previous section we found that the 2D Luttinger Hamiltonian can
be mapped exactly to a Hamiltonian of non-interacting nodal bosons
coupled to the antinodal fermions. This model is similar to the
standard model of metallic electrons coupled to phonons.  Similarly as
for the latter model one can eliminate the nodal bosons and thus
obtain an effective model for the antinodal fermions.

We derive this effective model using a standard functional integral
formalism \cite{NO}. To simplify our presentation in the main text we
work in position space and suppress the UV and IR cutoffs as in
Section~\ref{sec2}. All formulas below can be made precise in Fourier
space; see Appendix~\ref{appC1}.

We represent the partition function $\cZ$ of the bosonized 2D
Luttinger model in \Ref{Ha1}, \Ref{Hn11} and \Ref{Hna11} as functional
integral over the real-valued fields $\Phi_\pm=\Phi_\pm(\tau,\vx)$ and
Grassmann number valued fields $\psi_{\pm,0}=\psi_{\pm,0}(\tau,\vx)$
with $\tau\in[0,\beta)$ the usual Matsubara time and $\beta>0$ the
inverse temperature. Denoting the standard functional integral
measures as $D[\cdots]$ we obtain
\begin{equation*}
\cZ\define \textrm{Tr}(\ee^{-\beta H}) = \int
D[\Phi_+,\Phi_-]D[\psi_{+,0},\psi_{-,0}] \, \ee^{-S_a-S_n-S_{na}} 
\end{equation*}
with the actions $S_{(\cdot)}=\int_0^\beta d\tau\int d^2 x\,
\cL_{(\cdot)} $ where
\begin{equation*}
  \cL_n = \frac12\sum_{s=\pm} \left[ 
    \frac1{v_F(1-\gamma)}(\partial_\tau\Phi_s)^2 + 
    v_F(1+\gamma)(\partial_s\Phi_s)^2 \right] 
  + v_F\gamma(\partial_+\Phi_+)(\partial_-\Phi_-) 
\end{equation*} 
with $\partial_\tau=\frac{\partial}{\partial\tau}$,
\begin{equation}
  \cL_a = \sum_{r=\pm} \psi_{r,0}^\dag\left( \partial_\tau + 
    rc_F\partial_+\partial_-
    +c_F'(\partial_+^2+\partial_-^2) -\mu_0\right) \psi_{r,0}^{\phantom\dag} 
  + 2g_3  J_{+,0}J_{-,0} 
\end{equation} 
with $J_{\pm,0}=\psi^\dag_{\pm,0}\psi^{\phantom\dag}_{\pm,0}$, and
\begin{equation*} 
  \cL_{na} =
  (\pi\tilde{a})^{-1/2} g_4 \sum_{r,s=\pm} J_{r,0} \partial_s\Phi_s. 
\end{equation*}  
By standard abuse of notation we denote the integration variables in
the functional integral by the same symbol as the corresponding
operators in the Hamilton formalism, and we suppress the common
arguments $\tau,\vx$.

The boson integral is Gaussian and can be computed exactly, and the
result has the following form,
\begin{equation}
\label{Sa} 
\cZ=\cZ_n \int D[\psi_{+,0},\psi_{-,0}]\ee^{-S_{\mathrm{eff}}} 
,\quad 
\cZ_n=\int D[\Phi_+,\Phi_-] \, \ee^{-S_n} 
\end{equation} 
with $-\log\cZ_n/(\beta L^2)$ the nodal contribution to the free
energy and
\begin{equation}
  S_{\mathrm{eff}} = S_a + 
  \int_0^\beta d\tau\int d^2 x\int_0^\beta d\tau'\int d^2 y
  \sum_{r,r'=\pm} J_{0,r}(\tau,\vx) v_{\mathrm{eff}}(\tau-\tau',\vx-\vy) 
  J_{0,r'}(\tau',\vy)
\end{equation} 
the effective action for the antinodal fermions, including the
effective antinodal two-body potential $v_{\mathrm{eff}}$ induced by
the nodal bosons. We compute $v_{\mathrm{eff}}$ and $\cZ_n$ in
Appendix~\ref{appC1}.

We obtain the following Fourier transformed effective
potential\footnote{i.e.\ $v_{\mathrm{eff}}(\tau,\vx) = 1/(\beta
  L^2)\sum_{\omega,\vp}\hat{v}(\omega,\vp)
  \ee^{-\ii\omega\tau-\ii\vp\cdot\vx}$}
\begin{equation}
\label{veff} 
\hat{v}_{\mathrm{eff}}(\omega,\vp) = 
-\frac{(g_4)^2}{\pi\tilde{a}v_F}
\sum_{s=\pm} \frac{W_s(\vp)}{\omega^2 + \omega_s(\vp)^2} 
\end{equation}
with the functions 
\begin{equation}
\label{Ws} 
W_{s=\pm}(\vp) =  \begin{cases}   
    v_F^2(1-\gamma)
    \left(|\vp|^2+s\frac{(p_+^2-p_-^2)^2+\gamma|\vp|^4}{(1+\gamma)\sqrt{|\vp|^4 
          - A(2p_+p_-)^2}}\right) & 
    \mbox{if }\; 
    |p_\pm|\leq \pi\kappa_{\min}/(\sqrt{2}a)
    \\
    2v_F^2(1-\gamma)p_s^2 & 
    \mbox{if }\begin{cases}  
        |p_\pm|\leq \pi\kappa/(\sqrt{2}a)\; \mbox{ and}\\
        |p_\pm|\not\leq \pi\kappa_{\min}/(\sqrt{2}a)\ 
\end{cases} 
\\
    & \\[-1.5ex] 
    0 & \mbox{otherwise} 
\end{cases} 
\end{equation}
and $\omega_{s=\pm}(\vp)$ in \Ref{Es}; $\omega\in 2\pi\Z/\beta$ are
the usual boson Matsubara frequencies. Note that the second option in
\Ref{Ws} can only occur if $\kappa<1/2$ (since otherwise
$\kappa_{\min}=\kappa$), and similarly in \Ref{veff1} below.

As expected from the above-mentioned analogy with phonons, the
effective interaction induced by the nodal bosons has a non-trivial
time dependence.  One can study the effective antinodal fermion model
defined by the action in \Ref{Sa} using a generalized mean field
theory allowing for frequency dependent order parameters (this would
be similar to the Migdal-Eliashberg theory of electron-phonon systems
\cite{Mi,El}), but it is useful to have a simpler treatment. We
therefore approximate the time dependent effective potential by an
interaction local in time as follows,
\begin{equation}
\label{local} 
v_{\mathrm{eff}}(\tau,\vx)\approx \delta(\tau)\int_0^\beta
v_{\mathrm{eff}}(\tau',\vx)d\tau' = \delta(\tau)
\sum_{\vp\in\tilde\Lambda^*} \fLL \, \hat{v}_{\mathrm{eff}}(0,\vp)
\ee^{\ii\vp\cdot\vx}
\end{equation}
(arguments to justify this approximation can be found in the last
paragraph of this section and in Appendix~\ref{appC2}).  In this time
local approximation the effective antinodal action in \Ref{Sa} can be
obtained from the Hamiltonian
\begin{equation}
\label{Heff} 
H_{\textrm{eff}} = H_a + \sum_{\vp\in\tilde\Lambda^*} \fLL
\sum_{r,r'=\pm} \hat{v}_{\textrm{eff}}(0,\vp)
:\! \hat{J}_{r,0}(-\vp)\hat{J}_{r',0}(\vp)\!: 
\end{equation} 
with $H_a$ in \Ref{Ha} and the colons indicating normal ordering with
respect to the state $\Omega$.

We find that
\begin{equation}
\label{veff1}
\hat{v}_{\mathrm{eff}}(0,\vp)) = \begin{cases} 
    -g_{\textrm{eff}} & \mbox{if }\; 
    |p_\pm|\leq \pi\kappa_{\min}/(\sqrt{2}a)\\
    -g_{\textrm{eff}}\frac{1+\gamma/2}{1+\gamma} &    
    \mbox{if }\begin{cases} 
        |p_\pm|\leq \pi\kappa/(\sqrt{2}a)\; \mbox{ and}\\
        |p_\pm|\not\leq \pi\kappa_{\min}/(\sqrt{2}a) 
\end{cases}  \\
0& \mbox{otherwise} 
\end{cases} 
\end{equation}
with the constant $g_{\textrm{eff}}\define (g_4)^2/[\pi\tilde{a}
v_F(1+2\gamma)]$. Inserting \Ref{const}, \Ref{gj} and \Ref{tildea} we
obtain $g_{\textrm{eff}}$ in \Ref{geff}.  It is remarkable that
$\hat{v}_{\textrm{eff}}(0,\vp)$ is piecewise constant.

Since $\hat{J}_{\pm,0}(\vp)$ is non-zero only if $|p_\pm|\leq 
\pi\kappa/(\sqrt{2}a)$, and $\hat{v}_{\textrm{eff}}(0,\vp)$ is
constant and equal to $-g_{\textrm{eff}}$ in that region for
$\kappa\leq 1/2$, one can simplify the effective Hamiltonian in
\Ref{Heff} for $\kappa\leq 1/2$ as follows,
\begin{equation}
  \label{Heff1} 
  H_{\textrm{eff}} = H_a - 2g_{\mathrm{eff}} 
  \sum_{\vp\in\tilde\Lambda^*} \fLL
  \hat{J}_{+,0}(-\vp)\hat{J}_{-,0}(\vp)
\end{equation} 
(see Appendix~\ref{appC3} for details).  We thus obtain the remarkable
result that, for $\kappa\leq 1/2$, the only effect of the nodal bosons
in the time local approximation is a renormalization $g_3\to
g_3-g_{\textrm{eff}}$ of the antinodal coupling constant.

We note that $\pi\kappa_{\min}/(\sqrt{2}a)$ in \Ref{veff1} can be
regarded as a UV cutoff for the effective antinodal interaction, and
according to Hypothesis~H1 in Section~\ref{sec3.2} we expect that it
should be possible to remove this cutoff without significantly
changing the low energy physics of the model. This suggests that we
can simplify the effective antinodal model and use \Ref{Heff1} even
for $\kappa>1/2$.

To justify the approximation in \Ref{local} we note that it is similar
to the one used by Bardeen and Pines to derive the phonon induced
electron-electron interaction in a metal by eliminating the phonons in
an electron-phonon Hamiltonian using a similarity transformation
\cite{BP}.  Indeed, our bosonized 2D Luttinger Hamiltonian has the
very same structure as an electron-phonon Hamiltonian, and one can
therefore eliminate the nodal bosons in the Hamiltonian formalism
using the method in \cite{BP}. This yields
\begin{equation*}
  H^{\mathrm{BP}}_{\textrm{eff}} = H_a +  
  \sum_{\vp,\vk,\vk'} \fLL(\tPiL)^4
  \sum_{r,r'=\pm} \hat{v}_{rr'}(\vk,\vp)
  :\! 
  \hat\psi^\dag_{r,0}(\vk)\hat\psi^{\phantom\dag}_{r,0}(\vk+\vp)
  \hat\psi^\dag_{r',0}(\vk')\hat\psi^{\phantom\dag}_{r',0}(\vk'-\vp)
  \!: 
\end{equation*} 
with the following effective potential
\begin{equation*}
  \hat{v}_{rr'}(\vk,\vp) = -\frac{(g_4)^2}{\pi\tilde{a}v_F}
  \sum_{s=\pm}
  \frac{W_s(\vp)}{2\omega_s(\vp)}\left( 
    \frac1{\omega_s(\vp) - \Delta\EE_{r,0}(\vk,\vp)} + 
    \frac1{\omega_s(\vp) + \Delta\EE_{r',0}(\vk,\vp)}\right)   
\end{equation*}
depending on the energy differences $\Delta\EE_{r,0}(\vk,\vp)\define
\EE_{r,0}(\vk)-\EE_{r,0}(\vk-\vp)$.  However, different from
electron-phonon systems, in the present case these energy differences
are typically much smaller than $\omega_\pm(\vp)$ for $|\vk|,|\vk'|\ll
\pi/a$ (since $\EE_{\pm,0}(\vk)$ scales like $ta^2|\vk|^2$ and
$\omega_\pm(\vp)$ like $ta|\vp|$). Moreover, since $W_s(\vp)$ vanishes
like $(ta|\vp|)^2$ for $\vp\to \vzero$, this effective interaction
remains finite in this limit. This suggests that it is legitimate to
approximate this interaction by setting $\Delta\EE_{r,0}(\vk,\vp)=0$.
This yields our simplified time local approximation in \Ref{Heff}. In
the present case the fermion-boson interaction need not be small,
however, and thus the argument in \cite{BP} is not conclusive. We
therefore give a complementary argument to justify our approximation
in Appendix~\ref{appC2}.  It would be interesting to adapt
Migdal-Eliashberg theory \cite{Mi,El} to the present case and study
the effective antinodal action without such approximation.

\section{Final remarks}
\label{sec7}

\noindent\textbf{1.} As mentioned in the introduction, the  
nodal Hamiltonian in \Ref{Hn1} is essentially Mattis' model
\cite{Mattis}. The relevance of the antinodal Hamiltonian in \Ref{Ha1}
as an effective model for 2D lattice fermion systems was emphasized by
Schulz \cite{Schulz}. We argue in this paper that these two models
capture complementary aspects of the physics of 2D lattice fermion
systems, and our 2D Luttinger model includes them both.

\noindent\textbf{2.} Together with de Woul we studied the effective
antinodal Hamiltonian derived in Section~\ref{sec6.3} using mean field
theory \cite{dWL1}. We found a significant parameter regime away from
half-filling where the antinodal fermions are half filled
($\nu_a=1/2$) and fully gapped \cite{dWL1}. We expect that the
antinodal fermions do not affect low energy properties if they are
fully gapped. We therefore propose that the exactly solvable Mattis
model in \Ref{Hn1} is an appropriate low energy effective model in
this regime. 

\noindent\textbf{3.} There exists also a parameter regime
where the antinodal fermions are only partially gapped or not gapped
at all ($\nu_a\neq 1/2$) \cite{dWL1}. In this regime also the
antinodal fermions contribute to the low energy properties of the
system. We expect that this makes a qualitative difference in the
physical behavior, and that the mean field treatment of the antinodal
fermions is not as trustworthy for $\nu_a\neq 1/2$ as for $\nu_a=1/2$.

\noindent\textbf{4.} At very large doping values the condition 
in \Ref{condition} eventually fails.  In this regime we expect that
one can ignore all but the in- or out fermions ($s=2$ with $r=-$ or
$r=+$) for $\nu\ll 1$ or $1-\nu\ll 1$, respectively, and arguments
similar to the ones in Section~\ref{sec5} lead to the following
effective model for the 2D \ttpV model,
\begin{equation*}
  H=\sum_{\vk\in\Lambda^*_{\pm,2}} [(\mp c_F/2+c_F')\vk^2-\tilde\mu_{\pm,2}] 
  \hat\psi^\dag_{\pm,2}(\vk)  \hat\psi\pdag_{\pm,2}(\vk)   
\end{equation*} 
(the interaction vanishes in our Approximation~A2 due the the Pauli
principle).  The renormalized chemical potential $\tilde\mu_{2,\pm}$
is determined by filling $\nu$. This effective model is trivially
solvable. It suggests that the 2D \ttpV model describes a Fermi
liquid in the overdoped regime.

\noindent\textbf{5.} The 2D Luttinger models has three parameters more
than the 2D \ttpV model, namely $\kappa$, $Q$ and $\nu_a$. Our
approach becomes more predictive if one can determine these
parameters. As shown in \cite{dWL1}, one can compute $Q$ and $\nu_a$
self-consistently by using mean field theory for the antinodal
fermions. This leaves only one free parameter, namely $\kappa$. We
expect that $\kappa$ can be fixed by comparison with experimental data
or by the requirement that the mean field phase diagram of the
effective antinodal fermion Hamiltonian is consistent with the one of
the 2D \ttpV model. It would be interesting to fix $\kappa$ by a
variational computation.

\noindent\textbf{6.} Approximations~A1--A5 in Section~\ref{sec5} 
amount to adding or ignoring terms in the Hamiltonian that, as we
argued in Section~\ref{sec3.2}, are not relevant for the low energy
properties of the system. It would be interesting to investigate in
more detail if these approximations are relevant or not. This can be
done, in principle, using our results: one can also bosonize the terms
that were modified in our approximations (recall that the
approximations that do not affect the nodal fermions can be easily
avoided). It should be possible to investigate the relevance of these
terms using renormalization group methods, similarly as in \cite{SL}.

\noindent\textbf{7.} It is known that the truncated nodal 
model in \Ref{Hn} with $g_2=0$ has ``Luttinger liquid'' behavior after
\cite{VM} but not before Approximation~A5 \cite{ZYD}. We believe that
our model with $g_2>0$ is less sensitive to Approximation~A5 than the
truncated model with $g_2=0$. The reason is that the boson dispersion
relations in \Ref{Es} are quite isotropic, whereas for $g_2=0$ they
are $\omega_\pm(\vp)\propto |p_\pm|$ independent of $p_{\mp}$. We
therefore expect that the boson propagator has better decaying
properties in Fourier space for $g_2>0$ than for $g_2=0$.

\bigskip

\noindent {\bf Acknowledgments.} I would like to thank Alan Luther,
Vieri Mastropietro, Alexios Polychronakos, Manfred Salmhofer, Asle
Sudb{\o}, Mats Wallin, and in particular Jonas de Woul, for their
interest and useful discussions.  I am grateful to Jonas de Woul for
carefully reading the manuscript and suggesting various improvements,
and to Boris Fine and Manfred Salmhofer for helpful discussions on
BCS-like interaction terms. I also thank anonymous referees for
constructive criticism. This work was supported by the Swedish Science
Research Council~(VR), the G\"oran Gustafsson Foundation, and the
European Union through the FP6 Marie Curie RTN {\em ENIGMA} (Contract
number MRTN-CT-2004-5652).

\begin{appendix}

\section{Interaction terms}
\label{appA}
This appendix contains details of the computation mentioned in
Section~\ref{sec5.2}.

To find all solutions of \Ref{QQQQ} we collected the vectors
$\vQ_{r_1,s_1}-\vQ_{r_2,s_2}$ in a $8\times 8$ matrix with rows and
columns labeled by $(r_1,s_1)$ and $(r_2,s_2)$, respectively. We used
this matrix to find, for each case $(r_1,s_1,r_2,s_2)$, all cases
$(r_3,s_3,r_4,s_4)$ such that \Ref{QQQQ} holds true. For $Q=\pi/2$ we
found $8^3=512$ such cases. For $Q\neq\pi/2$ this number is reduced to
the 196 cases listed in table~\ref{table1}.

We now compute the interaction terms 3--11 in table~\ref{table1}. We
suppress sums over $r,s$ (and $r'$ and/or $s'$ if applicable) when
possible. Note that the possible values of $r,s$ (etc.) are different
for the different cases; see table~\ref{table1}.

The diagonal Hartree terms 3 are
\begin{equation*}
\VV_3 = \sum_{r,s} \sum_{\vk_j} (\tPiL)^6 \hat{u}(\vzero)
\delta_{\vk_1-\vk_2+\vk_3-\vk_4,\vzero}\,
  \hat\psi^\dag_{r,s}(\vk_1)
  \hat\psi^{\phantom\dag}_{r,s}(\vk_2)
  \hat\psi^\dag_{r,s}(\vk_3)
  \hat\psi^{\phantom\dag}_{r,s}(\vk_4).
\end{equation*}
Rewriting this by renaming $\vk_2,\vk_4$ to $\vk_4,\vk_2$, adding the
result to $\VV_3$ above, and dividing by $2$ yields
\begin{eqnarray*}
 \VV_3 = \sum_{r,s} \sum_{\vk_j} (\tPiL)^6 \hat{u}(\vzero) 
\delta_{\vk_1-\vk_2+\vk_3-\vk_4,\vzero}\, 
 \frac12 \Bigl(
    \hat\psi^\dag_{r,s}(\vk_1)
    \hat\psi^{\phantom\dag}_{r,s}(\vk_2)
    \hat\psi^\dag_{r,s}(\vk_3)
    \hat\psi^{\phantom\dag}_{r,s}(\vk_4) + \\ 
    \hat\psi^\dag_{r,s}(\vk_1)
    \hat\psi^{\phantom\dag}_{r,s}(\vk_4)
    \hat\psi^\dag_{r,s}(\vk_3)
    \hat\psi^{\phantom\dag}_{r,s}(\vk_2) 
  \Bigr). 
\end{eqnarray*} 
We now use the canonical anticommutator relations of the fermion field
operator twice and obtain the result in \Ref{V3}.

The sum of terms 4 and 5 can be written as
\begin{equation*}
\begin{split}
  \VV_{4+5} = \sum_{\vk_j,r} \left(\tPiL\right)^{6}
  u(\vQ_{r,s}-\vQ_{-r,s'})\delta_{\vk_1-\vk_2+\vk_3-\vk_4,\vzero}\,  
  \Bigl(
  \hat\psi^\dag_{r,s}(\vk_1)
  \hat\psi^{\phantom\dag}_{-r,s'}(\vk_2) 
  \hat\psi^\dag_{-r,s}(\vk_3)
  \hat\psi^{\phantom\dag}_{r,s'}(\vk_4) \\ + 
  \hat\psi^\dag_{r,s}(\vk_1)
  \hat\psi^{\phantom\dag}_{r,s'}(\vk_2) 
  \hat\psi^\dag_{-r,s}(\vk_3)
  \hat\psi^{\phantom\dag}_{-r,s'}(\vk_4) \Bigr) 
\end{split}
\end{equation*} 
since $\hat{u}(\vQ_{r,s}-\vQ_{-r,s'}) = \hat{u}(\vQ_{r,s}-\vQ_{r,s'})$
for all cases $(s,s')$ listed in table~\ref{table1}. Renaming in the
second term $\vk_2,\vk_4$ to $\vk_4,\vk_2$ and using the canonical
anticommutator relations of the fermion field operators twice we find
$\VV_{4+5}=0$.

The back-scattering terms 6 are
\begin{equation*}
\VV_6 = \sum_{\vk_j} \left(\tPiL\right)^{6}
\hat{u}(\vQ_{r,s}-\vQ_{-r,s})\delta_{\vk_1-\vk_2+\vk_3-\vk_4,\vzero}
\, \hat\psi^\dag_{r,s}(\vk_1) \hat\psi^{\phantom\dag}_{-r,s}(\vk_2)
\hat\psi^\dag_{r,s}(\vk_3) \hat\psi^{\phantom\dag}_{-r,s}(\vk_4). 
\end{equation*}
Renaming $\vk_1,\vk_3$ to $\vk_3,\vk_1$ and using the canonical
anticommutator relations we find that $\VV_6=-\VV_6$, i.e.\ $\VV_6=0$.

The BCS terms 7 and 9 are 
\begin{equation*}
  \VV_7 =
  \sum_{\vk_j} \left(\tPiL\right)^{6} 
  \hat{u}(\vQ_{r,s}-\vQ_{r',s'})\delta_{\vk_1-\vk_2+\vk_3-\vk_4,\vzero} \,
  \hat\psi^\dag_{r,s}(\vk_1)
  \hat\psi^{\phantom\dag}_{r',s'}(\vk_2) \hat\psi^\dag_{r,s}(\vk_3)
  \hat\psi^{\phantom\dag}_{-r',s'}(\vk_4)=0 
\end{equation*} 
and 
\begin{equation*}
  \VV_9 = 
  \sum_{\vk_j} \left(\tPiL\right)^{6}
  \hat{u}(\vQ_{r,s}-\vQ_{r',s'})
  \delta_{\vk_1-\vk_2+\vk_3-\vk_4,\vzero} \,
  \hat\psi^\dag_{r,s}(\vk_1)
  \hat\psi^{\phantom\dag}_{r',s'}(\vk_2) \hat\psi^\dag_{r,s}(\vk_3)
  \hat\psi^{\phantom\dag}_{r',s'}(\vk_4)=0
\end{equation*} 
by the same argument as for $\VV_6$ above. Since $\VV_8$ is the
Hilbert space adjoint of $\VV_7$ this also proves $\VV_8=0$.

For the terms 10 we find $\hat{u}(\vQ_{r,s}-\vQ_{r',s'})=0$ for all
cases listed in table~\ref{table1}, and thus $\VV_{10}=0$.  

Finally, the sum of all terms 11 is 
\begin{equation*}
\begin{split}
  \VV_{11} = \sum_{\vk_j}\sum_{r,r'} \left(\tPiL\right)^{6}
  \hat{u}(\vQ) \delta_{\vk_1-\vk_2+\vk_3-\vk_4,\vzero} \,\Bigl(
  \hat\psi^\dag_{r,0}(\vk_1) \hat\psi^{\phantom\dag}_{-r,0}(\vk_2)
  \hat\psi^\dag_{r',2}(\vk_3)
  \hat\psi^{\phantom\dag}_{-r',2}(\vk_4) + \\
  \hat\psi^\dag_{r,2}(\vk_1) \hat\psi^{\phantom\dag}_{-r,2}(\vk_2)
  \hat\psi^\dag_{r',0}(\vk_3) \hat\psi^{\phantom\dag}_{-r',0}(\vk_4)
  \Bigr).
\end{split} 
\end{equation*} 
Using the canonical anticommutator relations and inserting \Ref{hatu}
we obtain \Ref{V11}.

\section{Bosonization in 1D} 
\label{appB} 
We collect some well-known mathematical results that we need to
bosonize the nodal fermion Hamiltonian. More details can be found in
\cite{LM,CR,DS}, for example.

Consider the Hamiltonian
\begin{equation}
\label{H0L}
H_0 = \int_{-L/2}^{L/2} dx\, \sum_{r=\pm} \sum_{n=-M}^{M} rv_F
\!:\!\chi^{\dag}_{r,n}(x)
(-\ii \partial)\chi^{\phantom\dag}_{r,n}(x)\!: 
\end{equation} 
with $\chi^{(\dag)}_{r,n}(x)$ 1D fermion operators obeying
canonical anti-commutator relations normalized such that
\begin{equation*}
  \{\chi^{\phantom\dag}_{r,n}(x),\chi^{\dag}_{r',n'}(y)\} =
  \delta_{n,n'}\delta_{r,r'}\delta(x-y), 
\end{equation*}
$v_F>0$ a constant, $\partial=d/dx$, and $n$ a flavor index; $M>0$ is
some integer. We use anti-periodic boundary conditions, i.e.\ Fourier
transform is
\begin{equation*}
\hat\chi_{r,n}(k) = (2\pi)^{-1/2}\int_{-L/2}^{L/2} dx\,
\chi_{r,n}(x)\ee^{-\ii kx}\; 
\mbox{ with }\; k\in\tPiL(\Z+\half), 
\end{equation*} 
and the colons indicate normal ordering with respect to a
vacuum state $\Omega$ defined by the following conditions, 
\begin{equation*} 
\hat\chi^{\phantom\dag}_{r,n}(
k)\Omega=\hat\chi^{\dag}_{r,n}(-k)\Omega = 0 \quad \forall
rk>0 . 
\end{equation*} 
This Hamiltonian describes $2M+1$ flavors of 1D Dirac fermions. We also
introduce the fermion density operators
\begin{equation}
  j_{r,n}(x)= \; :\!\chi^{\dag}_{r,n}(x)\chi^{\phantom\dag}_{r,n}(x)\!:.   
\end{equation} 

We now state the well-known facts that allow to bosonize this
Hamiltonian (see e.g.\ \cite{CR} or \cite{DS} for proofs): Firstly, the
fermion densities obey the following non-trivial commutator relations,
\begin{equation} 
\label{R1}
[j_{r,n}(x),j_{r',n'}(y)] =
r\delta_{r,r'}\delta_{n,n'} \frac{1}{2\pi\ii}\partial\delta(x-y). 
\end{equation} 
Secondly, the Fourier transformed fermion densities 
\begin{equation*}
  \hat\jmath_{r,n}(p)=\int dx\, j_{r,n}(x)\ee^{-\ii px},\quad 
p\in\tPiL(\Z+\half)
\end{equation*} 
obey $\hat\jmath_{r,n}(p)^\dag=\hat\jmath_{r,n}(-p)$ and
\begin{equation}
\hat\jmath_{r,n}(p)\Omega = 0 \quad \forall rp\geq 0. 
\end{equation} 
Finally, 
\begin{equation}
\label{R2}
r\int_{-L/2}^{L/2} dx
:\!\chi^{\dag}_{r,n}(x)(-\ii \partial)
\chi^{\phantom\dag}_{r,n}(x)\!:\,    
= \pi \int_{-L/2}^{L/2} dx :\!j_{r,n}(x)^2 :\! . 
\end{equation}

We now observe that, by identifying $x_\pm$ with $x$, $x_{\mp}$ with
$n\tilde{a}$, $2M+1=L/\tilde{a}$, and
\begin{equation*}
  \chi_{r,n}(x) =  \sqrt{\tilde{a}}\psi_{r,\pm}(\vx), 
\end{equation*} 
the 1D Luttinger Hamiltonian in \Ref{H0L} is identical with the free
nodal Hamiltonian in \Ref{H0pm}. We thus can identify
\begin{equation*}
 j_{r,n}(x) = \tilde{a} J_{r,\pm}(\vx)
\end{equation*}
and write \Ref{R1} as
\begin{equation*} 
[J_{r,\pm}(\vx),J_{r',\pm}(\vy)] =
r\delta_{r,r'}\frac{1}{2\pi\ii\tilde{a}}\partial_\pm
\tilde\delta^2(\vx-\vy)
\end{equation*} 
with $\tilde\delta^2(\vx-\vy) = \delta(x_\pm-y_\pm)
\delta_{x_\mp,y_\mp}/\tilde{a}$.  Moreover, summing the identity in
\Ref{R2} over $n$ we get
\begin{equation*}
  \int d^2x\, 
  r:\! 
  \psi^\dag_{r,\pm}(\vx)(-\ii \partial_\pm)\psi_{r,\pm}^{\phantom\dag}(\vx)\!: = 
  \tilde{a}\pi \int d^2x\, :\! J_{r,\pm}(\vx)^2\!: 
\end{equation*} 
with $\int
d^2x=\int_{-L/2}^{L/2}dx_\pm\sum_{x_{\mp}\in\Lambda_{\mathrm{1D}}}\tilde{a}$.
The results summarized in the proposition in Section~\ref{sec6.1} are
obtained from this and other identities above by Fourier
transformation.

We finally note that our scaling is such that 
\begin{equation*} 
  \frac{\tilde{a}}{L} :\!N_{r,\pm}\!: \, =
  \frac{\tilde{a}}{L}\sum_{x_\mp}\tilde{a}\int dx_\pm J_{r,\pm}(\vx) =
  \frac1{2M+1} \sum_{n=-M}^{M} \int dx\, j_{r,n}(x)
\end{equation*} 
are well-defined operators in the quantum field theory limit $L/a\to\infty$,
and therefore
\begin{equation}
\label{Nn} 
\mbox{$(\frac{a}{L})^2$}\langle :\!N_{r,\pm}\!: \rangle \to 0\; \mbox{ as }\; 
L/a\to\infty. 
\end{equation}

\section{Effective interaction}
\label{appC} 
This section contains details on how the results presented in
Section~\ref{sec6} were obtained.

\subsection{Integrating out the nodal bosons}
\label{appC1}
To make precise the actions in the beginning of Section~\ref{sec6.3}
we work in Fourier space. We suppress the arguments and write
$\hat\Phi_\pm$ short for the Fourier transform
$\hat\Phi_\pm(\omega,\vp)$ of $\Phi_\pm(\tau,\vx)$, with $\omega\in
2\pi\Z/\beta$ the usual boson Matsubara frequencies. We also write
$\hat K_\pm$ short for
\begin{equation*}
  \hat K_{\pm}(\omega,\vp)=
    (\pi\tilde a)^{-1/2} 
g_4 \ii p_\pm\chi_\pm(\vp) \hat J_0(\omega,\vp)/(2\pi) 
\end{equation*} 
with $\hat J_0=\hat J_{+,0}+\hat J_{-,0}$.  Moreover,
$\hat\Phi_\pm^\dag(\omega,\vp)=\hat\Phi^{\phantom\dag}_\pm(-\omega,-\vp)$,
and similarly for $\hat K_\pm$ and $\hat J_0$.

Using \Ref{Hn2} and \Ref{Hna2} we can give the following precise
meaning to the action introduced at the beginning of
Section~\ref{sec6.3},
\begin{equation*}
\begin{split}
  S_n + S_{na} =\frac{2\pi^2}{\beta L^2} \sum_{\omega}\left(
    \sum_{s=\pm} \sum_{\vp\in\C_s}[
    A\pdag_s\hat\Phi^\dag_s\hat\Phi\pdag_s +
    2\hat{K}^\dag_s\hat{\Phi}\pdag_s] + \sum_{\vp\in\C}
    B[\hat\Phi^\dag_+\hat\Phi\pdag_-
    +\hat\Phi^\dag_-\hat\Phi\pdag_+]\right)
\end{split} 
\end{equation*} 
with 
\begin{equation*}
  A_{s=\pm}(\omega,\vp) \define 
  \frac{\omega^2}{v_F(1-\gamma\chi_s(\vp))} + 
  v_F(1+\gamma\chi_s(\vp))p_s^2 ,\quad
  B(\omega,\vp) \define v_F\gamma p_+p_-, 
\end{equation*} 
and 
\begin{equation*}
  \C \define\left\{\vp\in\tilde\Lambda^*:\; 
    |p_\pm|\leq \frac{\pi\min(\kappa,1-\kappa)}{\sqrt{2}a}\right\} 
\end{equation*}
the region in Fourier space where the $\hat\Phi_+$-$\hat\Phi_-$
coupling term in \Ref{Hn} is non-zero.

It is convenient to introduce the matrix notation 
\begin{equation*}
\underline{\hat\Phi}=\left(\begin{array}{c}\hat\Phi_+\\
    \hat\Phi_-\end{array}\right),\quad
\underline{\hat\Phi}^\dag=(\hat\Phi_+^\dag,\hat\Phi_-^\dag)
\end{equation*}
and similarly for $\hat K$. We also split the $\vp$-sums into sums
over $\C$ and the complements
\begin{equation*}
\C^\perp_s\define \left\{ \vp\in\C_s:\;
  \vp\notin \C\right\}.
\end{equation*} 
This allows us to write 
\begin{equation*}
  S_n+S_{na}  = \frac{2\pi^2}{\beta L^2} \sum_{\omega}
  \left( \sum_{\vp\in\C}\left[ \underline{\hat\Phi}^\dag 
      \underline{\underline{\cD}}^{-1} \underline{\Phi} +
      \underline{\hat\Phi}^\dag \underline{\hat{K}} +
      \underline{\hat{K}}^\dag \underline{\hat{\Phi}}
    \right]  + \sum_{s=\pm}\sum_{\vp\in\C_s^\perp} [
    A\pdag_s\hat\Phi^\dag_s\hat\Phi\pdag_s +
    2\hat{K}^\dag_s\hat\Phi\pdag_s]\right)  
\end{equation*} 
with
\begin{equation*}
\underline{\underline{\cD}}^{-1} = \left(\begin{array}{ll} 
A_+ & B \\
B & A_-
\end{array}\right). 
\end{equation*} 
We can now compute the nodal boson functional integral by completing
the square. This yields for the effective antinodal action defined in
\Ref{Sa}
\begin{equation*}
  S_{\mathrm{eff}} = S_a -  \frac{2\pi^2}{\beta L^2}\sum_{\omega}
  \left( \sum_{\vp\in\C} \underline{\hat{K}}^\dag
    \underline{\underline{\cD}}
    \underline{\hat{K}} + \sum_{s=\pm} 
    \sum_{\vp\in\C_s^\perp}\frac1{A_s} 
    \hat{K}^\dag_s\hat{K}\pdag_s \right)\equiv  S_a - \frac1\beta\sum_\omega 
  \sum_{\vp\in\tilde\Lambda^*} \fLL v_{\mathrm{eff}}\hat J_0^\dag\hat J\pdag_0
\end{equation*} 
with 
\begin{equation*}
  \hat{v}_{\textrm{eff}}(\omega,\vp) = 
  \begin{cases} 
      -(2\pi\tilde{a})^{-1} (g_4)^2
      (p_+,p_-)\, \underline{\underline{\cD}}(\omega,\vp)(p_+,p_-)^t & 
      \mbox{ if }\vp\in\C\\ 
      -(2\pi\tilde{a})^{-1}(g_4)^2 p_s^2/A_s(\omega,\vp) & \mbox{ if }
      \vp\in \C^\perp_s, s=+\;\mbox{ or }\; -\\
      0 & \mbox{ otherwise }
\end{cases}  
\end{equation*}
and ``$t$'' indicating matrix transposition. This implies the result
in \Ref{veff} \textit{ff}.  We also obtain
\begin{equation*}
  \cZ_n \propto \prod_{\omega} \prod_{s=\pm}
  \prod_{\vp\in\C_s}
  \frac{1}{\omega^2 + \omega_s(\vp)^2}  
\end{equation*} 
with $\omega_{s=\pm}(\vp)$ in \Ref{Es}. This expression for $\cZ_n$ is
formal but can be made well-defined by a multiplicative regularization
as follows,
\begin{equation}
\label{Zn} 
\cZ_n = \prod_{s=\pm}
\prod_{\vp\in\C_s}\prod_{n\in\Z} 
\frac{(2n\pi/\beta_0)^2+(v_Fp_s)^2}{(2n\pi/\beta)^2 + \omega_s(\vp)^2} 
= \prod_{s=\pm}\prod_{\vp\in\C_s}
\frac{\sinh^2(\beta_0 v_F|p_s|/2)}{\sinh^2(\beta \omega_s(\vp)/2)}
\end{equation} 
with $\beta_0>0$ arbitrary. This regularization amounts to multiplying
$\cZ_n$ with an irrelevant (infinite) constant. The nodal free energy
$-\log(\cZ_n)/(\beta L^2)$ computed from this generalizes the nodal
ground state energy $\cE_n$ in \Ref{cEn} to finite temperature.

\subsection{Locality of the effective interaction}
\label{appC2}
It is easy to see that $v_{\textrm{eff}}(\tau,\vx)$ decays like
$1/R^3$ for $R\define \sqrt{\vx^2+(v_F\tau)^2}\gg a$, different from
and electron-electron interaction potential induced by phonons which
decays like $1/R$ in 2D. This suggests that the former potential
decays quickly in time and that \Ref{local} is reasonable.

To see in more detail how this potential $v_{\textrm{eff}}(\tau,\vx)$
behaves for $L\gg |\vx|\gg a$ we compute it in the limits $a\to 0$ and
$L\to\infty$. Using \Ref{veff}, \Ref{Ws} and \Ref{Es} we obtain by
inverse Fourier transformation
\begin{equation*}
v_{\textrm{eff}}(\tau,\vx) = -g_{\textrm{eff}}\, \frac{1}{v_F^2|\tau|^3} 
f_\gamma\left(\varphi,\frac{|\vx|}{\sqrt{1-\gamma^2}\, v_F|\tau|}\right) 
\end{equation*}
with the special function
\begin{equation*}
f_\gamma(\varphi,x)=
\frac{(1+2\gamma)}{8\pi(1+\gamma)(1-\gamma^2)}\int_0^{2\pi}
\frac{d\chi}{2\pi} \sum_{s=\pm}
\frac{[1+sw(\chi)][e_s(\chi)^2 - 3
  x^2\cos^2(\chi-\varphi)]}{[e_s(\chi)^2 +
  x^2\cos^2(\chi-\varphi)]^3}
\end{equation*}
where
\begin{equation*}
  e_\pm(\chi) = \sqrt{\frac{1\pm\eta(\chi)}2},\quad w(\chi)
  =\frac{\cos^2(2\chi)+\gamma}{(1+\gamma)\eta(\chi)},\quad
  \eta(\chi)=\sqrt{1-A\sin^2(2\chi)}, 
\end{equation*}
$A=1-\gamma^2/(1+\gamma)^2$, and $(x_+,x_-) =
|\vx|(\cos\varphi,\sin\varphi)$. Plotting this function $f_\gamma(\varphi,x)$
using MATLAB we find that it does not change much with $\gamma$, that
it is positive for small values of $x$, and that it diverges like 
$1/|x|$ for $x\to 0$.  Moreover, for $x>\approx 1$ this function
oszillates strongly with $x$ and $\varphi$ and is predominantly
negative. Thus $v_{\textrm{eff}}(\tau,\vx)$ is sharply peaked and
strongly attractive for $|\tau|\ll |\vx|/v_F$, but for
$|\tau|>|\vx|/v_F$ it is strongly oszillatory.  The fact that $\int
d\tau\, v_{\textrm{eff}}(\tau,\vx)=-g_{\textrm{eff}}\, \delta^2(\vx)$
shows that this effective interaction averages in time to zero
everywhere except in $\vx=\vzero$. We thus expect that the local
approximation in \Ref{local} is reasonably accurate.

\subsection{Effective antinodal Hamiltonian}
\label{appC3}
This appendix contains details on the effective antinodal Hamiltonians
in \Ref{Heff} and \Ref{Heff1}.  We first explain why we have to normal
order in \Ref{Heff} with respect to $\Omega$, and then show how to
obtain \Ref{Heff1} from \Ref{Heff}.

In our computation in Section~\ref{sec6.3} we use a functional
integral formalism in which the antinodal fermions are represented by
Grassmann numbers. It is important that these Grassmann numbers are
defined using fermion coherent states with respect to the state $\Omega$
(see e.g.\ \cite{NO}) since this allows us to do the computation in a
way which is manifestly particle-hole symmetric. To convert the
Hamiltonian to a functional integral we therefore normal order with
respect to $\Omega$ and then replace the fermion operators by Grassmann
numbers dropping normal ordering. Similarly, to convert our result in
\Ref{Sa} to a Hamiltonian using the time local approximation in
\Ref{local}, we have to normal order the result with respect to the
state $\Omega$.

For $\kappa\leq 1/2$ the effective antinodal Hamiltonian in \Ref{Heff}
simplifies to
\begin{equation*}
H_{\textrm{eff}} = H_a - g_{\textrm{eff}} 
\sum_{\vp\in\tilde\Lambda^*} \fLL \sum_{r,r'=\pm} 
:\! \hat{J}_{r,0}(-\vp)\hat{J}_{r',0}(\vp)\!: . 
\end{equation*}
This differs from the result in \Ref{Heff1} by terms proportional to
\begin{equation*}
(*)_\pm \define \sum_{\vp}\fLL\, :\! \hat{J}_{\pm,0}(-\vp)
\hat{J}_{\pm,0}(\vp)\!:
\end{equation*} 
since $:\! \hat{J}_{\pm,0}(\vp)\!:\; = \hat{J}_{\pm,0}(\vp)$.  We now
show that $(*)_\pm=0$ by the Pauli exclusion principle.

We compute $(*)_\pm$ above by inserting $\hat{J}_{\pm,0}(\vp)=\; :\!
\rho_{\pm,0}(\vp)\!:$ and \Ref{hrho}. We obtain
\begin{equation*} 
  (*)_\pm=\sum_{\vk_j\in\Lambda^*_{\pm,0}} \fLL (\tPiL)^4
  \delta_{\vk_1-\vk_2+\vk_3-\vk_4,\vzero}:\!
  \hat\psi^\dag_{\pm,0}(\vk_1) \hat\psi^{\phantom\dag}_{\pm,0}(\vk_2)
  \hat\psi^\dag_{\pm,0}(\vk_3)
  \hat\psi^{\phantom\dag}_{0,\pm}(\vk_4)\!:  
\end{equation*} 
since $::\!A\!::\!B\!::\; = \; :\!AB\!:$.  We rewrite this by
renaming $\vk_2,\vk_4$ to $\vk_4,\vk_2$.  Under the normal ordering
sign fermion creation and annihilation operators can be anti-commuted,
and thus we find $(*)_\pm=-(*)_\pm$, i.e.\ $(*)_\pm=0$. This proves
our result in \Ref{Heff1}.

The argument above also shows that \Ref{Heff1} is obtained for
$\kappa>1/2$ provided that we approximate $\hat v_{\mathrm{eff}}$ by
removing the UV cutoff, as discussed at the end of
Section~\ref{sec6.3}.

\end{appendix}

\end{document}